\documentclass[acmsmall]{acmart}

\usepackage[T1]{fontenc}

\usepackage{multirow}
\usepackage{diagbox}
\usepackage{pifont}

\usepackage{tikz}
\newcommand*\emptycirc[1][.8ex]{\tikz\draw (0,0) circle (#1);} 
\newcommand*\halfcirc[1][.8ex]{%
	\begin{tikzpicture}
	\draw[fill] (0,0)-- (90:#1) arc (90:270:#1) -- cycle ;
	\draw (0,0) circle (#1);
	\end{tikzpicture}}
\newcommand*\fullcirc[1][.8ex]{\tikz\fill (0,0) circle (#1);}

\PassOptionsToPackage{table,xcdraw,dvipsnames}{xcolor}
\usepackage{xcolor}
\usepackage{xspace}
\definecolor{gree}{HTML}{1e8449}
\definecolor{orang}{HTML}{b35809}
\newcommand{\fish}[1]{\textcolor{black}{#1}\xspace}
\newcommand{\fishfix}[1]{\textcolor{gree}{#1}\xspace}

\usepackage{threeparttable}
\usepackage{colortbl}
\usepackage{hhline}
\usepackage{enumitem}
\usepackage{wrapfig}

\usepackage{array}

\newcolumntype{B}{>{\bfseries}c}
\newcolumntype{G}{>{\columncolor{gray!20}}c}

\newcommand{\worst}{\cellcolor{red!10}}
\newcommand{\best}{\cellcolor{green!10}}
\newcommand{\dataset}{CAShift\xspace}

\usepackage[most]{tcolorbox}
\usepackage{graphicx}
\usepackage{subcaption}

\AtBeginDocument{%
  }

\setcopyright{cc}
\setcctype{by}
\acmDOI{10.1145/3729346}
\acmYear{2025}
\acmJournal{PACMSE}
\acmVolume{2}
\acmNumber{FSE}
\acmArticle{FSE076}
\acmMonth{7}
\received{2024-08-26}
\received[accepted]{2025-04-01}

\begin{document}

\title{CAShift: Benchmarking Log-Based Cloud Attack Detection under Normality Shift}

\author{Jiongchi Yu}
\orcid{0000-0002-2888-4499}
\affiliation{%
  \institution{Singapore Management University}
  \city{Singapore}
  \country{Singapore}
}
\email{jcyu.2022@phdcs.smu.edu.sg}

\author{Xiaofei Xie}
\orcid{0000-0002-1288-6502}
\affiliation{%
  \institution{Singapore Management University}
  \city{Singapore}
  \country{Singapore}
}
\email{xfxie@smu.edu.sg}

\author{Qiang Hu}
\authornote{Corresponding Author.}
\orcid{0000-0002-8251-1669}
\affiliation{%
  \institution{Tianjin University}
  \city{Tianjin}
  \country{China}
}
\email{qianghu@tju.edu.cn}

\author{Bowen Zhang}
\orcid{0009-0009-7513-2319}
\affiliation{%
  \institution{Singapore Management University}
  \city{Singapore}
  \country{Singapore}
}
\email{bwzhang@smu.edu.sg}

\author{Ziming Zhao}
\orcid{0000-0003-1455-4330}
\affiliation{%
  \institution{Zhejiang University}
  \city{Hangzhou}
  \country{China}
}
\email{zhaoziming@zju.edu.cn}

\author{Yun Lin}
\orcid{0000-0001-8255-0118}
\affiliation{%
  \institution{Shanghai Jiao Tong University}
  \city{Shanghai}
  \country{China}
}
\email{lin_yun@sjtu.edu.cn}

\author{Lei Ma}
\orcid{0000-0002-8621-2420}
\affiliation{%
  \institution{University of Tokyo}
  \city{Tokyo}
  \country{Japan}
}
\affiliation{%
  \institution{University of Alberta}
  \city{Alberta}
  \country{Canada}
}
\email{ma.lei@acm.org}

\author{Ruitao Feng}
\orcid{0000-0001-9080-6865}
\affiliation{%
  \institution{Singapore Management University}
  \city{Singapore}
  \country{Singapore}
}
\affiliation{%
  \institution{Southern Cross University}
  \city{New South Wales}
  \country{Australia}
}
\email{ruitao.feng@scu.edu.au}

\author{Frank Liauw}
\orcid{0009-0009-1462-9794}
\affiliation{%
  \institution{Government Technology Agency of Singapore}
  \city{Singapore}
  \country{Singapore}
}
\email{Frank_LIAUW@tech.gov.sg}

\renewcommand{\shortauthors}{J. Yu, X. Xie, Q. Hu, B. Zhang, Z. Zhao, Y. Lin, L. Ma, R. Feng and F. Liauw}

\begin{abstract}
With the rapid advancement of cloud-native computing, securing cloud environments has become an important task. Log-based Anomaly Detection~(LAD) is the most representative technique used in different systems for attack detection and safety guarantee, where multiple LAD methods and relevant datasets have been proposed. However, even though some of these datasets are specifically prepared for cloud systems, they only cover limited cloud behaviors and lack information from a whole-system perspective. Another critical issue to consider is normality shift, which implies that the test distribution could differ from the training distribution and highly affect the performance of LAD. Unfortunately, existing works only focus on simple shift types such as chronological changes, while other cloud-specific shift types are ignored, e.g., different deployed cloud architectures. Therefore, a dataset that captures diverse cloud system behaviors and various types of normality shifts is essential.

To fill this gap, we construct a dataset \textit{CAShift} to evaluate the performance of LAD in cloud, which considers different roles of software in cloud systems, supports three real-world normality shift types~(application shift, version shift, and cloud architecture shift), and features 20 different attack scenarios in various cloud system components. Based on \textit{CAShift}, we conduct a comprehensive empirical study to investigate the effectiveness of existing LAD methods in normality shift scenarios. Additionally, to explore the feasibility of shift adaptation, we further investigate three continuous learning approaches, which are the most common methods to mitigate the impact of distribution shift. Results demonstrated that 1) all LAD methods suffer from normality shift where the performance drops up to 34\%, and 2) existing continuous learning methods are promising to address shift drawbacks, but the ratio of data used for model retraining and the selection of algorithms highly affect the shift adaptation, with an increase in the F1-Score of up to 27\%. Based on our findings, we offer valuable implications for future research in designing more robust LAD models and methods for LAD shift adaptation.

\end{abstract}

\ccsdesc[500]{Security and privacy~Intrusion/anomaly detection and malware mitigation}
\ccsdesc[500]{Software and its engineering~Contextual software domains}



\keywords{Cloud Native Systems, Software Vulnerabilities, Normality Shift, Anomaly Detection, Intrusion Detection, Log Analysis}



\maketitle

\section{Introduction}

Cloud-native infrastructures featured with Kubernetes~\cite{k8s} and Docker~\cite{docker} have rapidly emerged as the predominant choice for modern cloud computing architectures, subsequently becoming the trending foundational infrastructure for hosting numerous software systems such as Software as a Service (SaaS) platforms. Existing study reports that Kubernetes has a commanding \textbf{92\%} market share among container orchestration tools used by businesses throughout the world~\cite{datadog-study} and over 50\% of Fortune-100 companies have adopted Kubernetes~\cite{oracle-study}. Despite the success, various security-critical attacks are key factors limiting the reliable use of cloud systems, such as the severe cloud vulnerabilities CVE-2024-21626~\cite{cve-2024-21626} that can result in container escape due to mishandling of file descriptor isolation. Therefore, ensuring the security of cloud systems is crucial. 

Log-based Anomaly Detection (LAD) is a popular and effective method to guarantee the security of software-intensive systems. In real-world practice, system maintainers typically deploy LAD systems to monitor all tenant services for detecting threats within the cloud system that hosts various user applications. LAD systems collect and analyze system logs to train models that define normal operational behavior and identify anomalies.
Logs that significantly differ from the normality distribution are flagged as potential system anomalies or vulnerabilities exploited by attackers.

However, although many LAD methods~\cite{du2017deeplog,meng2019loganomaly,el2022contextualizing,yang2024try,zhao2021empirical} have been proposed with promising detection performance in many existing datasets~\cite{zhu2023loghub,grimmer2019modern}, few of them~\cite{el2022contextualizing,el2024replicawatcher} have focused on attacks in container clouds in a narrow scope (e.g., Docker). Besides, most of the proposed LAD methods are primarily evaluated using in-distribution test cases which follow the same data distribution as the training data. The continuous updating characteristics present in real-world software deployment environments, which result in changes to the data distribution of test data over time~(known as \textit{normality shift}~\cite{han2023anomaly}), affecting the performance of LAD methods, are overlooked. \fish{Cloud systems frequently undergo updates that can lead to shifts in system behaviors, which in turn affect the characteristics observed in system logs. Specifically, as software is continuously integrated, the features of normal logs collected to train normality models may experience shifts due to changes in software behavior. These changes include variations in function call types, frequencies, orders, and even noises of environmental components, as new software versions often behave differently compared to the previous version. As a result, the usefulness of existing LAD methods in such scenarios is unclear.} \fish{Commonly, system maintainers employ continuous learning periodically to adapt LAD systems to new log characteristics and mitigate the normality shift.}

Some LAD datasets have been constructed considering simple normality shift scenarios. For example, Anoshift~\cite{dragoi2022anoshift} divides the Kyoto-2006+ traffic dataset~\cite{song2011statistical} into three chronological orders and explores the performance of different LAD methods on shifted data. LogHub~\cite{zhu2023loghub,jiang2024loghub2} has released two versions of datasets to include more applications and logs collected at different times to consider distribution shift scenarios. However, directly applying these datasets to train LAD models for cloud systems is challenging. \ding{182} \textit{Unsuitable data source.} Network traffic and system logs, which are the major log types in existing datasets, are orthogonal log sources. Traffic datasets are not suitable for cloud system attack detection as characteristics apparent in system logs may not be detectable through pure network traffic analysis, while the reverse is not necessarily true. 
\ding{183} \textit{Limited attack scenarios.} Prior datasets focus on application-level logs and ignore system-level logging behaviors. Therefore, vulnerabilities~(attacks) in system-level~(e.g., kernel or container systems) will be overlooked when using these datasets to train LAD models and lead to security issues. \ding{184} \textit{Lack of the entire cloud system consideration.} In container environments, external systems are transparent to the container user, resulting in events occurring outside of the Open Container Initiative (OCI) runtime not being captured by the log monitoring of a container. This oversight results in such datasets~\cite{el2022contextualizing,grimmer2019modern} lacking a comprehensive examination of the entire cloud system service and, correspondingly, missing higher-level cloud system vulnerabilities in the test set, including those related to Container Runtime Interface (CRI) runtimes~\cite{CVE-2022-1708,CVE-2020-15257} and Kubernetes system vulnerabilities~\cite{CVE-2021-25742,CVE-2021-25743}. \ding{185} \textit{Insufficient shift types.} Existing works focus only on a monotonous shift type, e.g., time shift, and lack a systematic investigation and discussion on other real-world shift types specific to LAD application scenarios, such as cloud architecture shift.

To address the above challenges, in this work, we construct a normality shift-aware dataset specifically designed for cloud attack detection, named \dataset. Specifically, \dataset thoroughly considers the crucial role of cloud-native system components in system call logs and includes a broader range of software vulnerabilities and attack types. \dataset encompasses container runtime vulnerabilities and cloud system attacks, which are overlooked in existing datasets. \fish{Additionally, \dataset supports multiple types of real-world normality shift, which could significantly affect LAD performance. In total, \dataset consists of 27,000 normal and 2,500 attack system call traces, 4.5 billion log entries, and 20 different attack scenarios, requiring more than 90 hours of recording.} Based on our collected dataset, we conduct a comprehensive empirical study to investigate the effectiveness of existing LAD methods on cloud systems and normality shift scenarios. In addition to existing LAD methods~\cite{du2017deeplog,meng2019loganomaly,el2022contextualizing,yang2024try,zhao2021empirical}, we design semantic-aware embedding methods for system call logs and implement AE and VAE models for evaluation. Our study aims to answer the following research questions: 

\textbf{RQ1: How effective are LAD methods in detecting attacks in cloud systems?} We first explore whether existing LAD methods can detect attacks in cloud systems through system call logs under in-distribution scenarios. The results will \ding{182} answer if we can directly employ LAD methods proposed for other systems to cloud systems, and \ding{183} be used as baselines to demonstrate the influence of normality shift on LAD methods.

\textbf{RQ2: How effective are LAD methods under normality shift?} In this research question, we explore how well existing LAD methods handle normality shift. We evaluate the performance of LAD models trained with in-distribution data on our prepared three types of shift datasets for each LAD baseline. 

\textbf{RQ3: Can continuous learning methods help with LAD shift adaptation?} In real-world production environments, cloud systems can generate a large volume of logs in a short period, making it challenging to label all logs in shift scenarios as new training data for shift adaptation. To this end, existing works~\cite{feng2020deepgini,chen2023quote} utilize continuous learning algorithms to select the most appropriate samples to retrain models for shift adaptation. \fish{We explore the effectiveness of three representative continuous learning methods (\textit{DeepGini}, \textit{ZOL}, \textit{KM-ST}), along with a random selection baseline for evaluating LAD shift adaptation.}

\noindent
\textbf{Contributions.} In summary, this paper makes the following contributions:
\begin{itemize}[left=0.1cm]

  \item We introduce the first LAD dataset specifically designed for cloud systems named {CAShift}, which considers different software roles in cloud systems and supports multiple types of normality shift. Compared to existing LAD datasets, \dataset encompasses extensive cloud native systems data, together with detailed analysis of the cloud attacks collected.
  
  \item We conduct a comprehensive empirical study to investigate the capabilities of existing LAD methods in cloud systems under normality shift. We find that 1) existing LAD models all experience a performance decrease in shift scenarios, with an average reduction of 17\% in F1-Score. 2) Continuous learning methods can help enhance the detection capabilities of LAD models under shift conditions, achieving up to 27\% in F1-Score improvement.
  
  \item We provide in-depth discussions for future research directions on LAD based on our experimental results. To benefit future research, we make our dataset and source code publicly available~\cite{website}. 
\end{itemize}


\section{Background and Related Works}
\label{background}

In this section, we introduce the background related to this work, including attacks in cloud systems, LAD, distribution shift in LAD, and continuous learning for shift adaptation.

\begin{figure}[h]
    \centering
\includegraphics[width=1\linewidth]{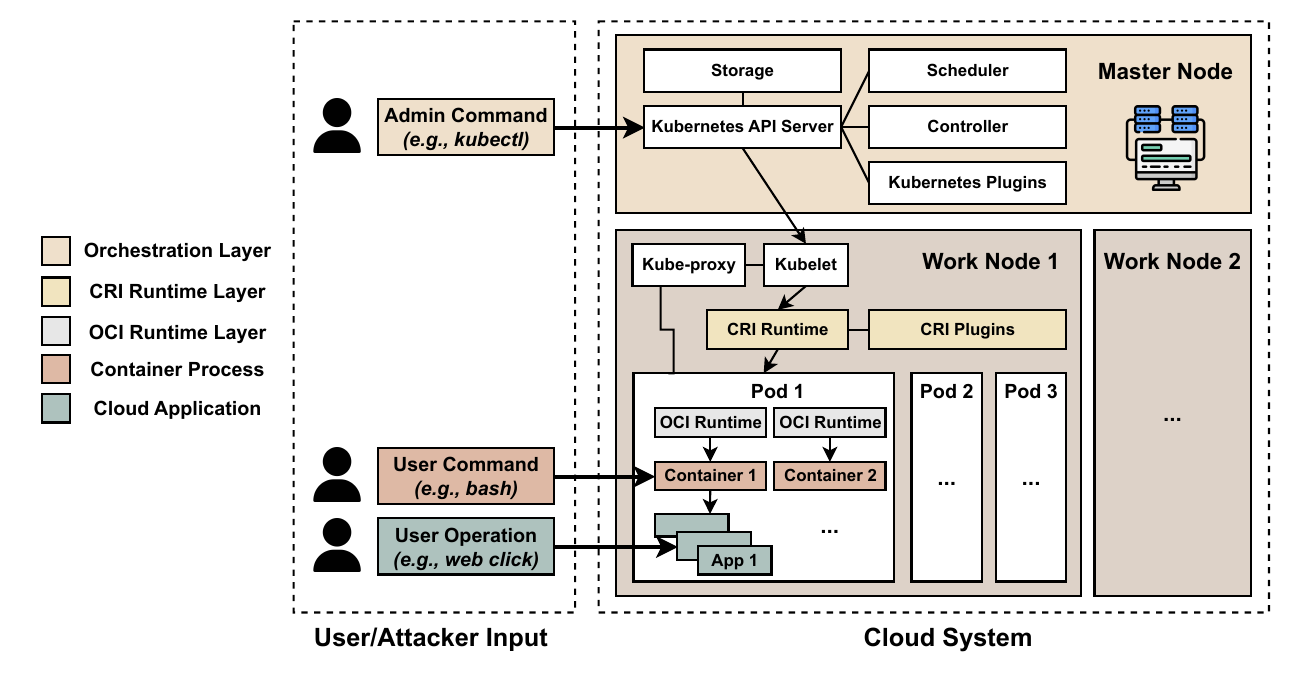}
    \Description{Common Attack Surfaces in Cloud Systems}
    \caption{Common attack surfaces in cloud systems.}
    \label{attacksurface}
\end{figure}

\subsection{Attacks in Cloud Systems} 
\label{section:problem definition}

Figure~\ref{attacksurface} illustrates common attack surfaces in cloud-native systems and their associated attack paths. A cloud system typically refers to a cloud orchestration framework deployed on physical machines, such as Kubernetes~\cite{k8s}. The cloud orchestration relies on cloud container runtimes~\cite{yu2024bug} to function, which can be subdivided into CRI runtimes~(e.g., containerd~\cite{containerd}) that manage images and container resources, and OCI runtimes~(e.g., runc~\cite{runc}) responsible for directly managing container instance processes and lifecycle. Cloud users usually interact with created containers, deploying various applications~(e.g., WordPress) to utilize cloud services. Despite the isolation provided by multiple \textit{cgroups} and \textit{namespaces} between the host system and user containers, user programs usually eventually run through the host kernel.

From an attacker's perspective, the entire cloud service system presents multiple potential exploitation entries. Specifically, vulnerabilities or misconfigurations in cloud applications deployed by users could allow attackers to gain command execution privilege within the application container~(e.g., Remote Code Injection vulnerabilities). Once the attacker can execute commands, vulnerabilities in the cloud container runtime can be exploited to escape to the upper cloud orchestration environment or even the host environment, including the popular runc escapes~\cite{CVE-2019-5736,cve-2024-21626}. In the host and cloud orchestration environments, vulnerabilities in the cloud system could be exploited to allow attackers to gain root privileges, including vulnerabilities in various cloud system components such as ingress~\cite{CVE-2021-25742} and the cloud system itself~\cite{CVE-2021-25743}. As attackers may employ different strategies at different stages of the attack process to acquire various resources, these activities vary significantly at the system call log level. This motivates us to construct a dataset that considers the collection of attack logs for vulnerabilities in different components and architectural components of cloud computing systems.

The goal of LAD for cloud systems is to distinguish normality logs and anomaly logs given a sequence of logs in cloud systems. A good LAD method should have three properties \ding{182} \textit{high accuracy}, which is the ability to precisely identify normal and attack logs, \ding{183} \textit{flexible adaptability}, which indicates the robustness of detecting performance in various cloud computing backgrounds, and \ding{184} \textit{scalability}, which refers to the capability to process the large volume of cloud system call logs without compromising on the speed or accuracy of threat detection. 



\subsection{Log-Based Anomaly Detection Methods}
\label{section:lad methods}

\fish{The task of LAD involves developing a model to predict whether incoming test logs are normal or anomalous. Traditional methods often leverage statistical techniques such as SemPCA~\cite{yang2024try} to transform high-dimensional log event features into a lower-dimensional subspace, enabling efficient distinguishing between normal and anomalous patterns. Advanced methods such as deep learning-based methods~\cite{zhao2021empirical,le2022log,zhao2024trident,zhao2023cmd} typically train the model to detect unknown threats that deviate from normal behaviors. These approaches can be further divided into two categories based on the representation of model output:}

\noindent
\textbf{Reconstruction-Based Method.} During training, this type of LAD methods reconstructs the input logs with the model and compares them with the initial input, aiming to minimize the reconstruction error of normal logs. At the test time, given a log sample, if the reconstruction loss exceeds a preset detection threshold, the log will be considered anomalous. Therefore, generative models such as AutoEncoder (AE)~\cite{el2022contextualizing} and Generative Adversarial Networks (GAN)~\cite{zenati2018adversarially} are frequently used as backbone models for these methods. Specifically, CHIDS~\cite{el2022contextualizing} processes system call log sequences into \textit{syscall sequence graphs} and uses the frequency of each system call as metrics for AE to learn from. To incorporate the semantic information of system calls, we implemented a standard AE model for LAD, which involves initially embedding all log information using a BERT~\cite{devlin2018bert} tokenizer. Subsequently, the class token of the embedded log is used as a learning input for the AE. Considering the different distributions in the LAD scenario, we also implement the Variational AutoEncoder~(VAE)~\cite{kingma2013auto} model as a baseline LAD method, which integrates the Kullback-Leibler~(KL) divergence between the hidden layers and the initial input embedded log into the reconstruction learning loss to make the learned data distribution more robust. 

\noindent
\textbf{Prediction-Based Methods.} \fish{These methods learn to maximize the probability of predicting the next log in normal log sequences~\cite{du2017deeplog,le2021log,zhao2024towards}. Consequently, the higher the normality prediction value for a log, which is the output of the model, the more likely it is considered normal. Conversely, if the prediction value falls below a preset detection threshold, it will be predicted as anomalous. Many unsupervised models have been proposed, including DeepLog~\cite{du2017deeplog}, LogAnomaly~\cite{meng2019loganomaly}, and LogAD~\cite{zhao2021empirical}. Among them, sequential DL models such as Long Short-Term Memory (LSTM), are often used as backbone models. Specifically, DeepLog utilizes the LSTM model to learn normal operations by predicting the next log event based on preceding events. Meanwhile, LogAnomaly builds on DeepLog by designing a synonyms-based method to represent log templates as semantic vectors, which introduces quantitative information to enhance anomaly detection. LogAD incorporates knowledge bases to integrate operational expertise in log preprocessing and apply ensemble prediction algorithms for different potential anomaly patterns, including keyword discrepancies and variable distributions.} \fish{Existing research, such as LogRobust~\cite{zhang2019robust}, also adopts supervised learning methods which require a large amount of labeled data to train models. Consequently, semi-supervised methods are proposed to utilize a subset of the training set containing labeled normal log data and cluster the unlabeled logs for LAD. For example, PLELog~\cite{yang2021semi} uses the GloVe global vector-based pre-trained language model~\cite{pennington2014glove} to represent each word in log events and constructs the model using an attention-based bidirectional GRU~\cite{cho2014properties} to predict the incoming logs.}

\subsection{Distribution Shift in LAD}

Distribution shift~\cite{koh2021wilds}, which indicates the test distribution is different from the training distribution, is a common problem in machine learning. In computer vision~\cite{hendrycks2019benchmarking,liu2024comprehensive}, natural language processing~\cite{koh2021wilds,gardner2024benchmarking}, and code learning~\cite{hu2023codes,hu2024active} fields, distribution shift has been widely studied to explore the generalization ability of machine learning models. In the field of LAD, as discussed in existing works~\cite{dragoi2022anoshift,hu2023codes,han2023anomaly}, distribution shift issues also harm the reliable usage of existing LAD methods in real-world scenarios. As time progresses, new normality features may differ from features learned by LAD models, leading to the distribution drift problem. Such drift behavior frequently happens in cloud and container systems due to their designed principle of ``build once, run anywhere" which allows minor updates to be continually implemented across cloud systems. These normality shifts can negatively impact to performance of trained LAD models.

Although most previous datasets have overlooked the issue of distribution shifts within their collection paradigms, typically maintaining a single release without considering the distribution shifts introduced over time, some datasets have acknowledged such problems and made improvements to their collected dataset. For instance, Anoshift~\cite{dragoi2022anoshift} categorizes the yearly updated data from the Kyoto-2006+ dataset~\cite{song2011statistical} into three chronological shift patterns to evaluate the performance of network traffic-based LAD. LogHub has also expanded its collected datasets with more covered applications in LogPub~\cite{jiang2024loghub2} and provided new versions for some of the sub-datasets with new collected time versions and pre-processed templates (e.g., Hadoop, HDFS). Different from the above works, we design three novel normality shift types and construct our dataset accordingly.  

\subsection{Continuous Learning for Shift Adaptation} 

Continuous learning is a process for machine learning models to continuously learn from new data and update their knowledge base. This approach allows models to adapt to new circumstances or data, thus serving as effective methods~\cite{hu2023codes,feng2020deepgini,chen2023quote,li2022hybridrepair} for enhancing model performance with new data distribution. Specifically, continuous learning involves two steps: 1) collecting new normality logs in shift distributions, and 2) retraining the model using the collected cases. \fish{In practice, system operators routinely update the LAD models to adapt emerging log patterns~\cite{han2023anomaly}. However, as the new log samples are normally unlabeled and could contain attacks, data annotation is a necessary step. Unfortunately, labeling log data is time-consuming and labor-intensive. Therefore, in this work, we consider employing label-efficient continuous learning methods to help shift adaptation for LAD methods. Four methods, namely \textit{DeepGini}, \textit{Zero-Order Loss} (ZOL), \textit{K-Multisection Strategy} (KM-ST), and \textit{Random} are included in our study.} \textit{DeepGini~\cite{feng2020deepgini}} prioritizes test cases based on the \textit{Gini impurity} of model outputs and selects cases that exhibit higher uncertainty for the model retraining process. \textit{ZOL} utilizes loss information to select test cases that are more likely to result in misclassification. To address the imbalanced loss distribution issue in \textit{ZOL}, which could lead to unfair selection of test cases through pure sequential selection, \textit{KM-ST}~\cite{chen2023quote} divides test cases into several distinct groups to ensure a balanced and comprehensive coverage across different input features. \fish{We also incorporate the \textit{Random Selection} method, which randomly selects logs for continuous learning as a baseline.} These methodologies are employed to select a certain proportion of shift logs for LAD model retraining, thereby facilitating adaptation to shift scenarios.



\section{CAShift: Benchmarking for Cloud Attack Detection}

\begin{figure}[t]
    \centering
    \includegraphics[width=1\linewidth]{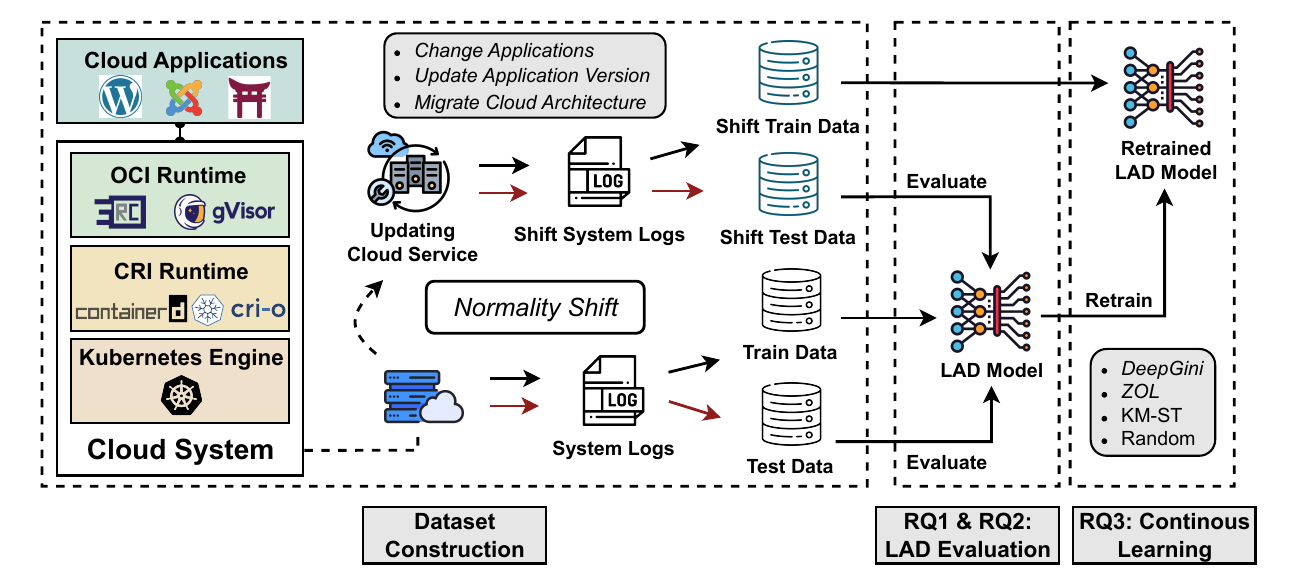}
    \Description{Overview of Benchmarking Framework}
    \caption{Overview of our benchmarking framework.}
    \label{fig:overview}
\end{figure}

In this section, we introduce the construction process of our benchmark datasets, which includes the data collection process and the quantitative analysis of shift scenarios. The overview of our collection framework is shown in Figure~\ref{fig:overview}.

\subsection{Normality Shift}

In total, three types of shift scenarios in cloud environments have been considered in CAShfit: \textit{Application Shift}, \textit{Version Shift}, and \textit{Cloud Architecture Shift}. 

\begin{itemize}[left=0.1cm]

\item \textit{Application Shift} comes from the different behaviors of different applications on the cloud. In practice, users often deploy multiple applications that are functionally different on the cloud side over time to meet their business needs. 
The logical changes within the applications can lead to variations in the invoking system functions, which introduce the application shift. 

\item \textit{Version Shift} is introduced by the new behaviors that come with new versions of software. Major version updates to many applications can lead to significant changes in functionality, which in turn affect the logic and sequence of system functionality usage. 

\item \textit{Cloud Architecture Shift} is a cloud system-specific shift type. Both CRI runtimes and OCI runtimes are regulated with their respective protocols, which have various implementations. For instance, \textit{containerd} and \textit{cri-o} are all CRI runtimes, and \textit{runc}, \textit{gvisor}, \textit{kata-containers}, and \textit{crun} all adhere to the OCI runtime specification~\cite{oci}. These runtimes exhibit significant differences at the system call level due to their internal characteristics in handling system calls for containers. Specifically, sandbox-based runtimes like \textit{gvisor} and \textit{kata-containers} implement system call wrappers for security concerns, whereas runtimes like \textit{cri-o} generate many system call logs inherent to their runtime operation. Consequently, different choices in cloud architecture can lead to varied behaviors in cloud system call logs. 

\end{itemize}

\subsection{Dataset Collection}

For the data construction, we follow existing works~\cite{el2022contextualizing,lin2020cdl} and collect system logs from three highly influential open-source applications (\textit{WordPress}, \textit{Joomla}, and \textit{Jinja}). In addition to the normal logs, \dataset contains attack logs based on existing CVE vulnerabilities found in these three applications. \fish{Considering shift scenarios, we collect logs from \textit{WordPress} and \textit{Joomla} under their three different versions and three different cloud container runtime environments.} We also include 20 types of CVE vulnerabilities from various components of cloud-native systems. We replay these vulnerabilities in their respective affected components and versions to collect the corresponding logs. After that, the capability of LAD methods in handling normality shift is evaluated by the collected datasets. Furthermore, continuous learning methods are employed to select important data to adapt LAD models to new data distributions.

As discussed in Section~\ref{background}, we focus on evaluating LAD for detecting cloud attack/anomaly behaviors in cloud systems under normality shift scenarios. Based on our shift definition, we construct our dataset CAShift, which supports three types of shift. Following previous works~\cite{grimmer2019modern,el2022contextualizing}, we select Kubernetes~\cite{k8s} as the base engine for the entire cloud system and adopt Sysdig~\cite{sysdig} to collect system call logs.  For \textit{application shift},  we deploy three applications \textit{WordPress}, \textit{Joomla}, and \textit{Jinja}, on cloud systems and capture both normal and attack logs from the system for our dataset. Specifically, we collect normal behaviors and gather 10 different vulnerabilities triggered by internal bugs in these applications, which are summarized in Table~\ref {table:Collected Vulnerability}, to replay for attack logs. For the \textit{version shift}, we collect logs of \textit{WordPress} in three stable versions including 4.8~(released on 08 Jun 2017), 5.6~(released on 08 Dec 2020), 6.2.1~(released on  29 Mar 2023), and \textit{Joomla} in versions 3.7, 4.2.7 and 5.1. \fish{For the \textit{cloud architecture shift}, we deploy \textit{WordPress} and \textit{Joomla} on different container runtime systems and collect logs accordingly.} Three most popular container runtime combinations, \textit{containerd (CRI) - runc (OCI)}, \textit{containerd (CRI) - gvisor (OCI)} and \textit{cri-o (CRI) - runc (OCI)} are considered in CAShift. \fish{We also assemble 10 different system-level vulnerabilities (e.g., Kubernetes, container runtimes, kernel) to enrich the diversity of \dataset in detecting vulnerabilities across various components in cloud computing environments. In total, CAShift consists of 27,000 normal and 2,000 attack system call traces with 4.5 billion log entries, and 20 different attack scenarios (with CVEs), assembling to more than 90 hours of recording.}

\noindent
\textbf{Normal Log Collection.} To collect normal behaviors of cloud systems, we follow previous works~\cite{grimmer2019modern,li2022enjoy} and utilize Selenium~\cite{selenium} to simulate normal user interactions on deployed cloud applications. Specifically, we simulate all possible user actions, including random link walks, comment submissions, and user panel logins~(non-administrative). The authors manually test the application and design different automated simulation schemes based on the application's functionalities. The system call logs are then collected as the normal behavior of cloud systems. 

\begin{table}[!t]
\centering
\caption{Collected vulnerability dataset.}
\label{table:Collected Vulnerability}
\resizebox{1\textwidth}{!}{
\begin{tabular}{|c|c|c|c|c|c|}
\hline
\textbf{Cloud Attacks} & \textbf{CVE/CWE ID} & \textbf{Vulnerability Info} & \textbf{CVSS} & \textbf{Bug Application} & \textbf{Bug Version} \\ \hline
\multirow{10}{*}{Application Level} & CWE-400 & Denial of Service (DoS) & - & WordPress & < 5.3 \\ \cline{2-6}
        & CVE-2016-10033 & Remote Code Execution (RCE) & 9.8 & WordPress & < 5.2.18 \\ \cline{2-6}
        & CVE-2019-17671 & Unauthorized Private Content Access & 5.3 & WordPress & < 5.2.4 \\ \cline{2-6}
        & CVE-2017-5487 & Unauthorized User Enumeration & 5.3 & WordPress & 4.7.1 \\ \cline{2-6}
        & CVE-2016-4029 & Server-Side Request Forgery (SSRF) & 8.6 & WordPress & < 4.5 \\ \cline{2-6}
        & CVE-2023-23752 & Improper Access Check & 5.3 & Joomla & 4.0.0 - 4.2.7 \\ \cline{2-6}
        & CVE-2021-23132 & Directory Traversal RCE & 7.5 & Joomla & 3.0.0 - 3.9.24 \\ \cline{2-6}
        & CVE-2017-8917 & SQL injection & 9.8 & Joomla & < 3.7.1 \\ \cline{2-6}
        & CVE-2015-8562 & Object Injection RCE & 7.5 & Joomla & < 3.4.6 \\ \cline{2-6}
        & CVE-2019-8341 & Server Side Template Injection (SSTI) & 9.8 & Jinja2 & 2.10 \\ \hline
\multirow{10}{*}{System Level} & CVE-2019-5736 & runC Overwrite Container Breakout & 8.6 & runC & 1.0-rc6 \\ \cline{2-6}
        & CVE-2021-30465 & Race-Based Container Breakout & 8.5 & runC & < 1.0.0-rc95 \\ \cline{2-6}
        & CVE-2024-21626 & Internal File Descriptor Leak & 8.6 & runC & < 1.1.11 \\ \cline{2-6}
        & CVE-2020-15257 & Privilege Escalation via API Socket & 5.2 & containerd & < 1.3.9 \\ \cline{2-6}
        & CVE-2022-1708 & Memory or Disk Space Exhaustion & 7.5 & cri-o & < 1.19.7 \\ \cline{2-6}
        & CVE-2020-14386 & Privilege Escalation via Memory Corruption & 7.8 & Kernel (gVisor Sentry) & < 5.9-rc4 \\ \cline{2-6}
        & CVE-2024-1086 & Use After Free (UAF) Privilege Escalation & 7.8 & Kernel & 5.14 - 6.6 \\ \cline{2-6}
        & CVE-2021-25742 & Unauthorized Cluster Secrets Stealth & 7.6 & Kubernetes Ingress Nginx & v1.0.0 \\ \cline{2-6}
        & CVE-2021-25743 & Insecure ANSI Escape Characters Filtering & 3.0 & Kubernetes & < 1.26.0-alpha.3 \\ \cline{2-6}
        & CWE-200 & Internal Service Spoofing & - & Kubernetes & 1.25 \\ \hline
\end{tabular}
}
\end{table}

\noindent
\textbf{Attack Log Collection.} \fish{The cloud attack vulnerabilities collected in \dataset are categorized into two major types: cloud application vulnerabilities and cloud system vulnerabilities. We follow a unified workflow to gather the attack behaviors. Specifically, for application attacks, we 1) explore vulnerabilities within the selected applications (i.e., WordPress, Joomla, Jinja2) from the MITRE CVE Database [20] and initially gather 11,505, 1,325, and 23 CVEs, respectively. 2) We manually review these CVEs and filter out dependencies or plugin-related CVEs, leaving 96 CVEs.
3) We rank the CVEs by their CVSS scores to identify the most impactful vulnerabilities, while also considering diversity in our selection process. For each distinct CWE category, we select only one representative attack. Consequently, we assemble 10 application-level CVEs in our dataset. A similar workflow is applied to collect 10 cloud attacks regarding identified cloud system components (i.e., Kubernetes, container, cri-o, runC, gvisor, Linux kernel).}

The detailed information of the selected vulnerabilities is shown in Table~\ref {table:Collected Vulnerability}. Notably, \dataset differs from existing datasets~\cite{grimmer2019modern,el2022contextualizing} as it considers vulnerabilities in various cloud computing components, which are often overlooked from a purely application or system attack perspective. \dataset includes unique cloud-specific attacks such as the newly discovered container escape vulnerability CVE-2024-21626~\cite{cve-2024-21626} and malicious service spoofing on Kubernetes clusters based on CWE-200~\cite{cwe-200}. After preparing the attacks, we replay every attack Proof of Concept (PoC)
in the bug version of the cloud components and collect the exploit system call logs. All attacks in \dataset are conducted without system admin privileges, closely mirroring real-world application scenarios and demonstrating the significant impact of these attacks. Overall, \dataset not only includes PoCs and the collected system logs of the vulnerability but also provides detailed vulnerability analysis (e.g., symptoms and root causes), offering insights for refined log analysis and future research purposes.

\begin{table}[!t]
\centering
\caption{Comparison of existing LAD datasets.}
\label{table: dataset}
\resizebox{1\textwidth}{!}{
\begin{threeparttable}
\begin{tabular}{cccccccccc}
\hline
\textbf{Dataset} & \textbf{Anomaly Scenario} & \textbf{Collection Base} & \textbf{Label Coarse} &\textbf{Application Shift} & \textbf{Version Shift} & \textbf{System Arch Shift} & \textbf{Normality Versatility} & \textbf{Log Complexity} & \textbf{Anomaly Diversity} \\ \hline
AnoShift~\cite{dragoi2022anoshift} & System Attack & NIDS Traffic Logs & per entry & $\times$ & $\checkmark$ & $\times$ & \halfcirc & \emptycirc & \halfcirc  \\ \hline
LANL-CMSCSE~\cite{akent-2015-enterprise-data} & System Attack & System Traffic Logs & per entry & $\times$ & $\checkmark$ & $\times$ & \halfcirc & \halfcirc & \emptycirc \\ \hline
LID-DS~\cite{grimmer2019modern} & Application Attack & System Call Logs & per window & $\checkmark$ & $\times$ & $\times$ & \fullcirc & \emptycirc & \halfcirc \\ \hline
CDL~\cite{lin2020cdl} & Application Attack & System Call Logs & per window & $\checkmark$ & $\times$ & $\times$ & \fullcirc & \halfcirc & \halfcirc \\ \hline
LogHub~\cite{zhu2023loghub,jiang2024loghub2} & System Anomaly & System Information & per entry & $\checkmark$ & $\checkmark$ & $\times$ & \fullcirc & \emptycirc & \fullcirc \\ \hline
CB-DS~\cite{el2022contextualizing} & System Attack & System Call Logs & per window & $\checkmark$ & $\times$ & $\times$ & \emptycirc & \halfcirc & \emptycirc \\ \hline
CAShift (Ours) & System Attack & System Call Logs & per window & $\checkmark$ & $\checkmark$ & $\checkmark$ & \fullcirc & \fullcirc & \fullcirc \\ \hline
\end{tabular}

\begin{tablenotes}
            \item \textbf{\fullcirc :True, \halfcirc :Partially True, \emptycirc :False} 
        \end{tablenotes}
    \end{threeparttable}
    }
\end{table}

\noindent
\textbf{Comparison of Existing Datasets.} Table~\ref{table: dataset} compares \dataset with existing datasets. \textit{Normality Versatility} indicates the variety of software sources covered by datasets. For example, CB-DS only collects logs from one self-implemented online shopping system, while CAShift includes logs sourced from different applications and distributed systems.  \textit{Log Complexity} refers to whether logs contain explicit or structured information that is easy for human interpretation, which indirectly represents the difficulty of the dataset for LAD methods. For instance, existing datasets contain many system anomalies that are always coupled with keywords including "warning" or "error", which is a distinct pattern for fine-grained log annotation and learning, and is easily detected by LAD methods. Conversely, attack information in CAShift often appears similar to normal system calls, which makes it more challenging to detect. \textit{Anomaly Diversity} represents the diversity of the collected anomaly scenarios of the dataset and considers whether the attacks are from different attack surfaces of the dataset domain. \fish{For example, CB-DS only contains four CVEs and six misconfiguration scenarios on container service, which limits the practical usage of this dataset. However, datasets like LogHub and \dataset consider various anomalies in the whole system view and support diverse scenarios. These three metrics combined with different shift types effectively demonstrate the diversity and learning challenges of the LAD dataset.} 




\subsection{Quantitative Analysis of CAShift}


In this part, we quantitatively analyze how data samples are distributed in CAShift by log embedding and token frequency visualization. 

\begin{wrapfigure}{r}{0.48\textwidth}
 \centering
 \includegraphics[height=0.48\textwidth]{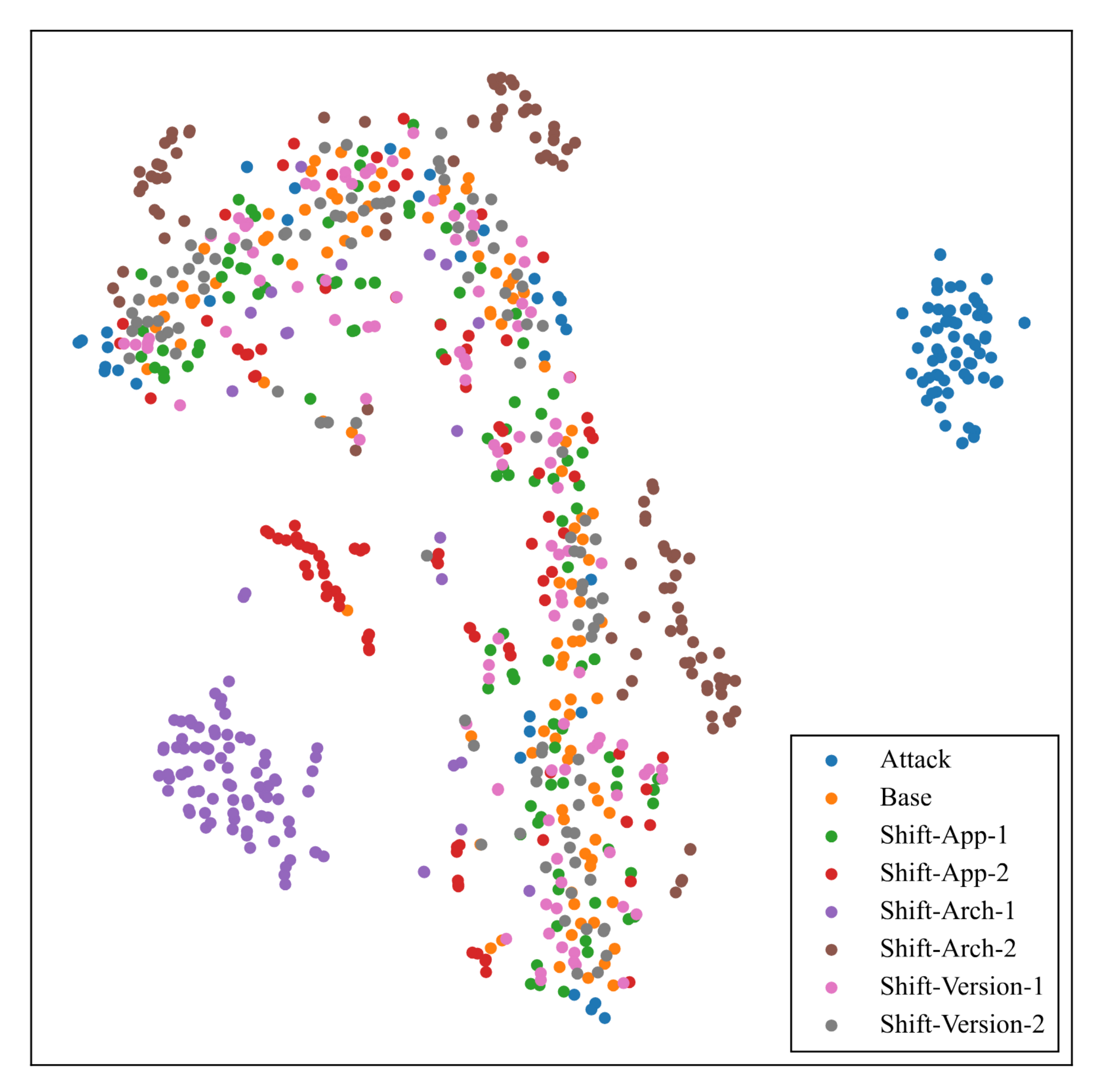}
 \caption{T-SNE visualization of shift logs compared to attack logs and normal logs.}\label{fig: tsne}
\end{wrapfigure}

\noindent
\textbf{T-SNE Analysis.} \fish{We employ the commonly used unsupervised method t-SNE~\cite{van2008visualizing} to help illustrate the embedding distribution of log information. Specifically, we randomly sample 100 logs from each shift scenario and all attack logs first and then employ the pre-trained BERT-base model to produce the embeddings of each log sample. Subsequently, we apply t-SNE to reduce the dimensionality of these embeddings to two dimensions for visualization purposes. Figure~\ref{fig: tsne} depicts the results, where different colors represent distinct groups of log data distributions, including base logs (i.e., Wordpress-6-containerd-runc) and various shift logs.}

\fish{We observe that different types of shifts deviate to varying degrees from the base distribution, which is considered as logs without shifts. Notably, cloud architecture shifts (e.g., Arch-1) show the most significant deviations compared to application shifts and version shifts. The results of the t-SNE visualization confirm our intuition that the analyzed cloud system logs are continuously shifting in different types and emphasize the need for a comprehensive study of existing LAD methods in detecting such shifted log data.}

\begin{figure}[h]
\centering

\begin{subfigure}[b]{0.48\textwidth}
    \centering
    \includegraphics[width=\textwidth]{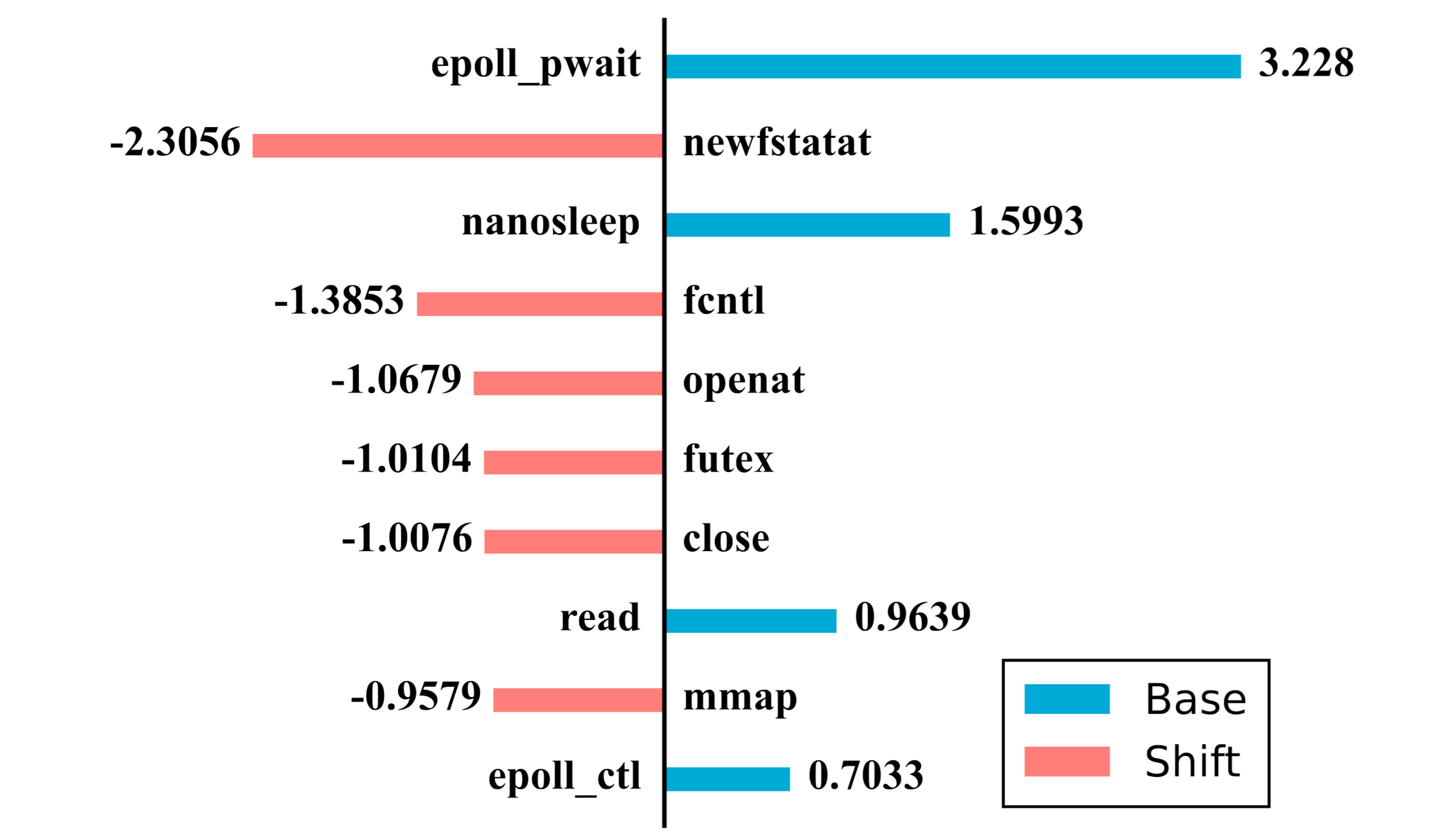}
    \caption{Application shift compared to Base}
    \label{fig:app1-shift}
\end{subfigure}
\hfill
\begin{subfigure}[b]{0.48\textwidth}
    \centering
    \includegraphics[width=\textwidth]{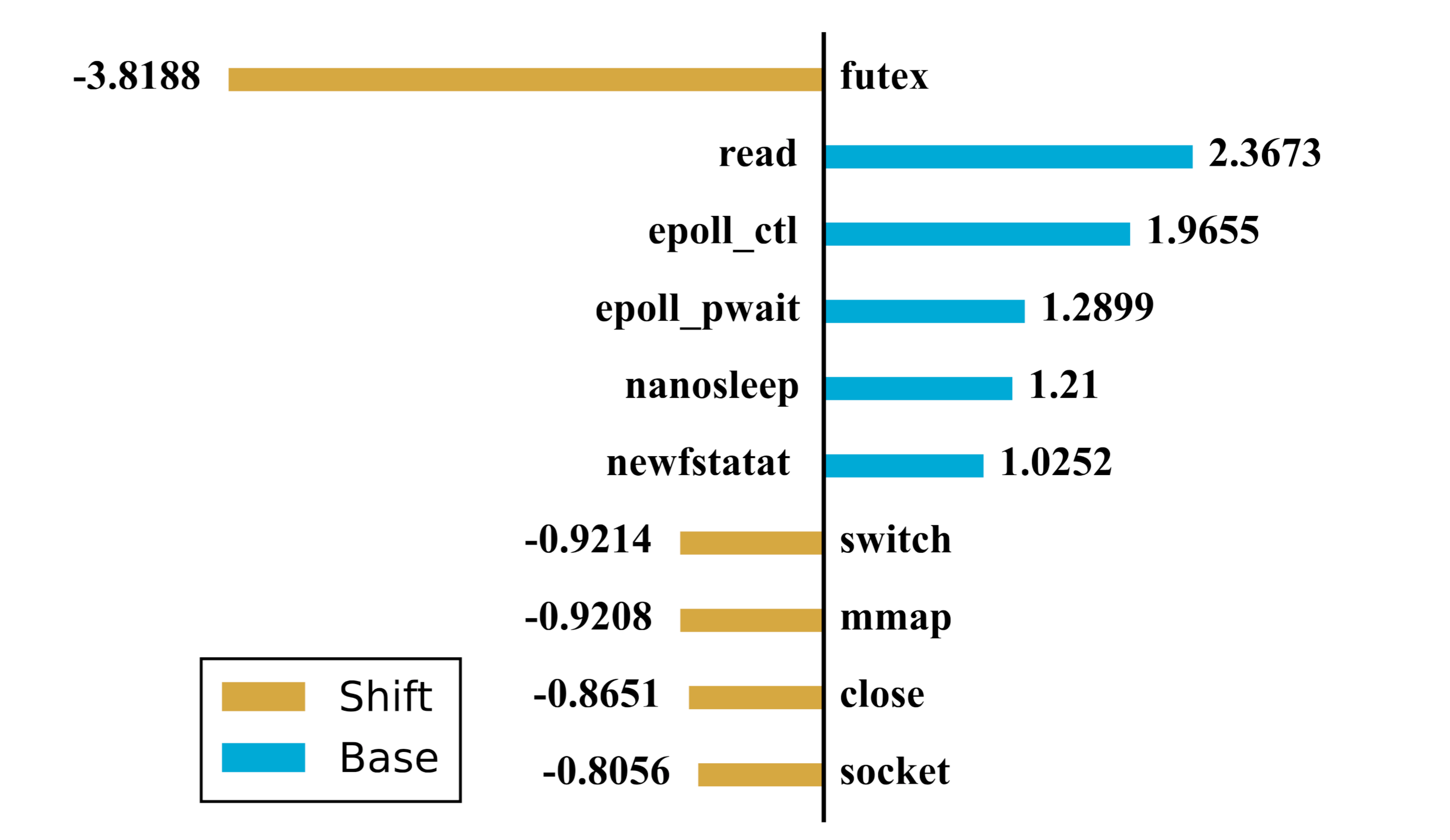}
    \caption{Version shift compared to Base}
    \label{fig:version1-shift}
\end{subfigure}

\begin{subfigure}[b]{0.48\textwidth}
    \centering
    \includegraphics[width=\textwidth]{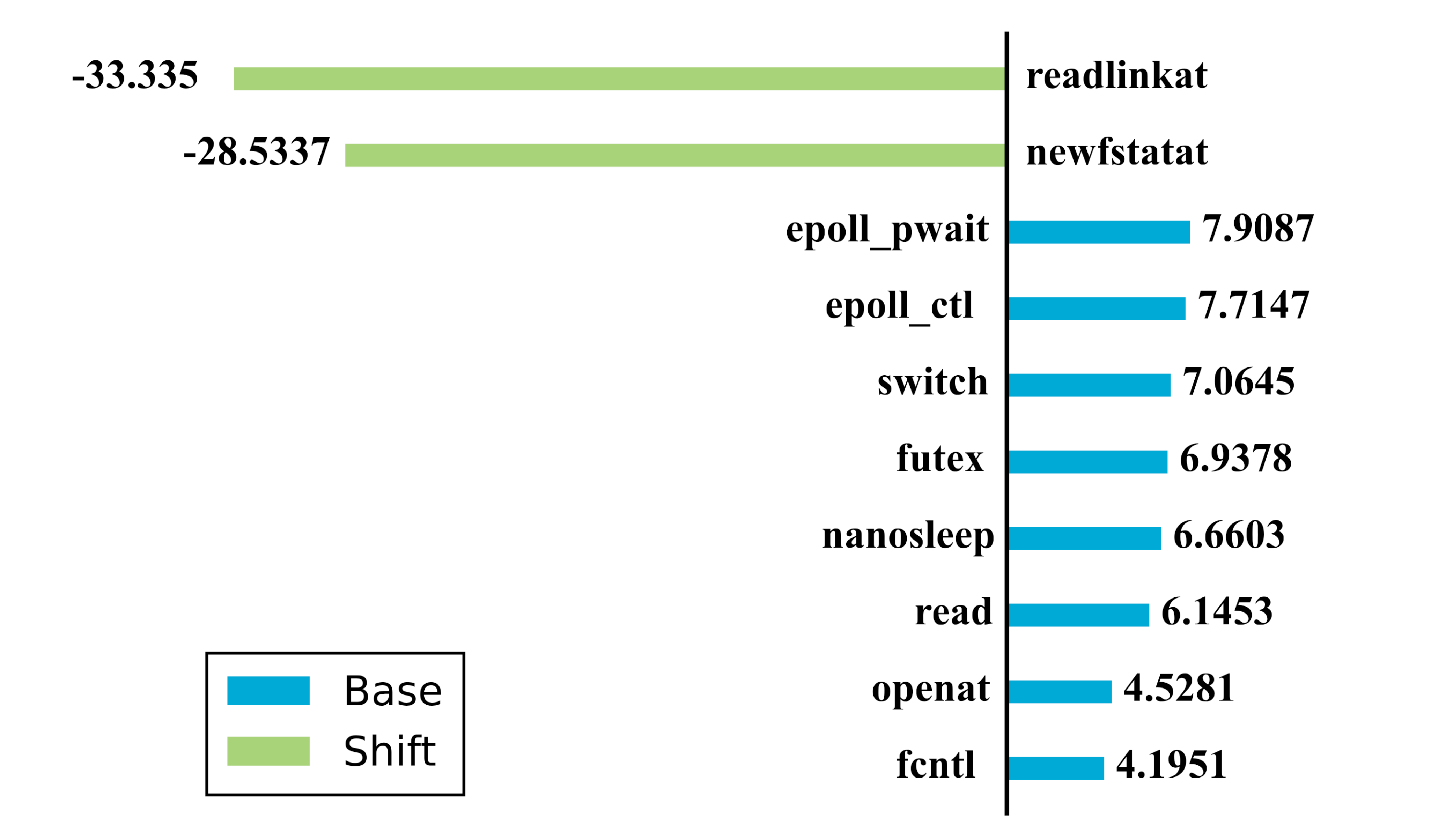}
    \caption{Cloud architecture shift compared to Base}
    \label{fig:arch1-shift}
\end{subfigure}
\hfill
\begin{subfigure}[b]{0.48\textwidth}
    \centering
    \includegraphics[width=\textwidth]{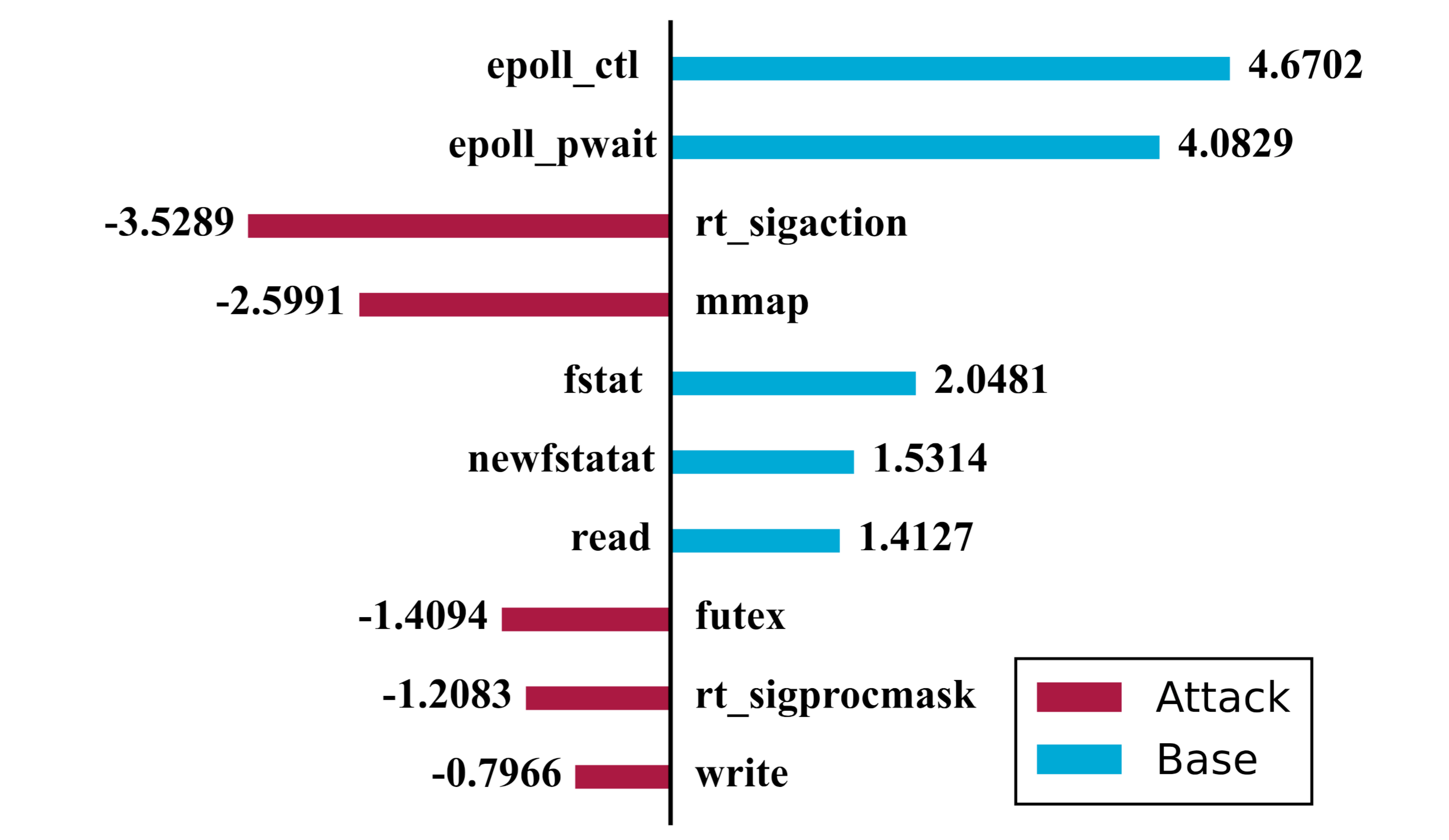}
    \caption{Attack logs compared to Normal}
    \label{fig:attack-base}
\end{subfigure}

\caption{The difference of the frequency~(\%) of top 10 system call names between shift distributions and the base distribution.}
\label{fig:statistic analysis}
\Description{Statistical Analysis for Shift Dataset}
\end{figure}

\noindent
\textbf{Log Statistical Characteristics Analysis.} \fish{Figure~\ref{fig:statistic analysis} depicts the system call frequency of logs in each distribution. Similar to natural language processing, LAD models treat logs as a sequence of tokens (e.g., system call names) and then produce the embedding. To this end, we extract all system call names from the collected log distributions and analyze the frequency of each system call's occurrence. Subsequently, we examine the changes in token frequency under different distribution shifts, and then rank these variations from highest to lowest, identifying the top ten system calls with the most significant distribution differences. The positive value suggests more system calls appeared in the shift scenarios, while the negative value indicates more system calls appear in the base distribution logs. We observe a clear token frequency shift between the shifted log distributions and the base logs. For the \textit{application shift} and \textit{version shift}, the biggest gap can be more than 3\%, which is close to the biggest gap between attack logs and normal logs~(4.67\%), indicating the significance of the shift. Surprisingly, for the \textit{cloud architecture shift}, the difference is even greater and can be up to more than 30\%, demonstrating that the cloud architecture changes introduce huge shift behaviors in the logs and potential influences regarding the performance of LAD models.}

\section{Empirical Study}
\label{design}

\subsection{Overview}

Based on \dataset, we conduct an extensive empirical study to investigate the effectiveness of LAD methods in detecting cloud attacks and their performance under three shifted scenarios. Furthermore, we explore whether existing continuous learning approaches can help LAD mitigate the influence distribution shift. 

\begin{itemize}[leftmargin=*]
    \item \textit{Data preparation.} First, based on the definition of normality shift, all logs in \dataset can be divided into multiple groups, e.g., for the application shift, logs collected from the same application are located in the same group. Then, we treat all the logs in one group as in-distribution data and logs from other groups as shifted data. Finally, we split the in-distribution and shifted datasets into training data and test data.

    \item \textit{LAD model training.} After the data preparation, we train LAD models using the in-distribution data and save the best model using validation data randomly split from the training data. Mention that, in addition to existing LAD methods~\cite{du2017deeplog,meng2019loganomaly,el2022contextualizing,chen2021experience}, we further design semantic-aware embedding methods for system call logs and implement VAE models as one of the baselines. 

    \item \textit{LAD model evaluation.} We then evaluate the trained LAD model using the in-distribution test data to explore how effective existing LAD methods are in detecting cloud system attacks to answer RQ1. Meanwhile, we use shifted test data to investigate the effectiveness of LAD methods in handling distribution shifts to answer RQ2.

    \item \textit{Shift adaption.} Finally, we employ continuous learning-based shift adaptation methods to select data from the shifted training dataset to label and retrain the trained LAD models in the last step accordingly. We then evaluate the performance of the retrained LAD models on shifted test data again to check whether these methods are useful in LAD scenarios and answer RQ3.

\end{itemize}

\subsection{Study Setup}

\fish{\textbf{Dataset configuration.} We use all our collected logs in the experiment. For the normal data, in each data distribution, including base logs and six shift scenarios. We randomly allocate 10\% of the logs as test data, which comprises approximately 150 system call traces. Each trace contains about 50 thousand log entries, featuring system logging information including timestamp, CPU ID, thread ID, system call name, and corresponding parameters. The remaining 90\% of the logs are divided into training and validation data at a 4:1 ratio. For the attack data, across all scenarios, a total of 200 attack system call traces selected from all affected vulnerabilities are used as the test dataset.}

\noindent
\textbf{LAD models.} Our evaluation considers six existing LAD methods including \textit{DeepLog}, \textit{LogAnomaly}, \textit{CHIDS}, \textit{SemPCA}, \textit{PLELog} and \textit{LogAD}. 
Since \textit{VAE}~\cite{kingma2013auto} has proved to be an effective embedding method to distinguish different data distributions, we employ it as one baseline in our study. We use publicly available implementations~\cite{le2022log,chids,yang2021semi,yang2024try} for the selected baselines and adhere to their reported configurations. In cases without official implementation, we implement based on the provided descriptions in the corresponding papers. We also design semantic-aware
embedding methods for the \textit{AE}~\cite{farzad2020unsupervised} and \textit{VAE} model and implement as additional baselines.

For \textit{AE} models, we utilize BERT tokenizer~\cite{devlin2018bert} to tokenize system call logs including timestamp, cpuID, system call, and corresponding arguments first. Then, these tokens are embedded in a vector-level representation with 768 dimensions. Finally, we utilize class tokens (CLS) of each embedded sequence to represent each system call log file and calculate the average reconstruction loss during the test phase. The log data is empirically encoded with a chunk size of 80 log entries with the token size 512~(which is the size limitation of the Bert base model).

The core design of \textit{VAE} lies in mapping data to a continuous latent space via its encoder and reconstructing data from this latent space through its decoder. Different from the \textit{AE model}, the training objectives of \textit{VAE} include minimizing reconstruction error to ensure the reconstructed data $\hat{x}$ closely matches the original log $x$ together with regularizing the latent space, typically by approximating the latent variables' distribution to a standard normal distribution. This regularization ensures the smoothness of the latent space, facilitating the adaptation of the model to new samples from different distributions. The loss used in our evaluation is calculated as follows, where $N$ represents the total number of input logs:

$$
\mathcal{L}_{\text{VAE}} = \sum_{i=1}^N [\text{MSE}(x_{i}, \hat{x}_{i}) + D_{KL}(q_\phi(z_{i}|x_{i}) \| \mathcal{N}(0,1))]
$$

In addition to calculating the Mean Squared Error (MSE) between the reconstructed log $\hat{x}$ and the initial input log $x$, VAE extracts the vector $z$ from the latent layer and computes the conditional probability distribution of the latent variable given the input $x$. The Kullback-Leibler (KL) divergence from the conditional probability distribution $q_\phi(z|x)$ to the standard normal distribution $\mathcal{N}(0,1)$ is then used in the loss function $\mathcal{L}_{\text{VAE}}$ to ensure that the distribution of the latent variables approximates the predetermined prior distribution. 

\noindent
\textbf{Continuous Learning Methods.} Following previous researches~\cite{hu2023codes,dang2023graphprior}, we adopt the popular continuous learning algorithms~\cite{feng2020deepgini,chen2023quote} to enhance LAD performance~(RQ3), including \textit{DeepGini}, \textit{ZOL}, \textit{KM-ST}, and \textit{Random}. We select the LAD models with the best performance result in shift environments~(RQ2) for evaluation, namely \textit{AE} and \textit{VAE}. We examine different proportions of logs with three algorithms introduced in these methods and use the selected logs to retrain the model. The selected continuous learning methods do not require modifications to the original model itself. Instead, logs are selected based on the reconstruction loss during the decoding process and the vectors in the latent layer. We load the pre-trained model and apply the same hyperparameters as the training process to retrain the model. 

\noindent
\textbf{Evaluation Metrics} We measure the performance of LAD methods using three metrics, \textit{Receiver Operating Characteristic - Area Under Curve (ROC-AUC)}, \textit{Precision}, \textit{Recall}, and \textit{F1-score}. Specifically, \textit{ROC-AUC} summarizes the classifier's performance across all possible classification thresholds and is independent of the model's threshold. \textit{ROC-AUC} approaches 1 indicates a stronger ability of the model to distinguish between data classes, while the \textit{ROC-AUC} value closer to 0.5 suggests the model cannot classify categories in the data. \textit{Precision} represents the proportion of true anomalies among all results reported as anomalies. \textit{Recall} measures the percentage of actual anomalies that were successfully identified as positive anomalies, and \textit{F1-Score} considers both the \textit{Precision} and \textit{Recall} of a classification model and serves as their harmonic mean. A higher F1-Score indicates better performance of the classifier. The calculation of each metric is summarized as follows: 

$$
Precision = \frac{TP}{TP+FP} \qquad Recall = \frac{TP}{TP+FN} \qquad F1-Score = \frac{2 \times Precision \times Recall}{Precision + Recall}
$$

True Positive~(TP) refers to anomaly logs that are correctly labeled as anomalies, while False Positive~(FP) is normal logs incorrectly labeled as anomalies. 

\noindent
\fish{\textbf{Reliability of Experiment Results.} To mitigate the effect of randomness introduced by the model training process and improve the reliability of the results, we apply a 10-fold cross-validation method to all experiments. We also conduct statistical significance analysis to examine the statistical significance of differences between LAD behaviors under the in-distribution and six-shift log data. For each evaluation scenario, we explore various hyperparameters to determine the optimal performance of LAD under different conditions. Due to space limitations, we report only the best results for each scenario in the paper.}

\section{Evaluation Result}
\label{result}

\subsection{RQ1: Effectiveness of LAD Methods under In-Distribution Scenarios}
\label{sec:rq1}

First, we investigate the effectiveness of LAD methods in cloud systems without considering distribution shifts. Table~\ref{table:rq1-baseline} presents the results of each method.  We can see that all existing LAD methods (e.g., \textit{DeepLog}, \textit{LogAnomaly}) demonstrate effective detection capabilities for cloud attacks with an average F1-Score over 0.9. This indicates that current methods can handle in-distribution scenarios. Our implemented VAE achieved the best results, reaching detection rates of approximately 0.99 for each attack.  However, some attacks remain challenging for some LAD baselines and can raise security issues. Specifically, prediction-based methods (including \textit{DeepLog} and \textit{LogAnomaly}) generally yield lower results for Denial of Service attacks within cloud applications, achieving F1-Scores of 0.8230 and 0.8663, respectively. We conjecture that DoS attacks inherently carry less semantic information compared to other attacks and the system calls of DoS attacks are often similar to normal ones~\cite{zlomislic2017denial}.  

\begin{table}[h]
\centering
\caption{F1-Scores of LAD methods in detecting different cloud attacks. The best and worst results are highlighted by the green and red background, respectively.}
\label{table:rq1-baseline}
\resizebox{1.0\textwidth}{!}{
\begin{tabular}{B|>{\centering\arraybackslash}m{2cm}|>{\centering\arraybackslash}m{2cm}|>{\centering\arraybackslash}m{2cm}|>{\centering\arraybackslash}m{2cm}|>{\centering\arraybackslash}m{2cm}|>{\centering\arraybackslash}m{2cm}|>{\centering\arraybackslash}m{2cm}|>{\centering\arraybackslash}m{2cm}}
\hline
\multirow{2}{*}{\diagbox{\textbf{Attacks}}{\textbf{Baselines}}} & \multirow{2}{*}{\textbf{DeepLog}} & \multirow{2}{*}{\textbf{LogAnomaly}} & \multirow{2}{*}{\textbf{CHIDS}} & \multirow{2}{*}{\textbf{PLELog}} & \multirow{2}{*}{\textbf{SemPCA}} & \multirow{2}{*}{\textbf{LogAD}} & \multirow{2}{*}{\textbf{AE}} & \multirow{2}{*}{\textbf{VAE}} \\ & & & & & & & & \\ \hline
\textbf{CWE-400}  & 0.8230 & 0.8663& \worst 0.8032 & 0.9277& 0.8950& 0.8395 & 0.9128  & \best 1.0000 \\ \hline
\textbf{CVE-2016-10033}  & 0.9091 & 0.9347& \worst 0.7843 & 0.9558& 0.9499& 0.9120 & 0.8867  & \best 0.9950   \\ \hline
\textbf{CVE-2019-17671}  & 0.8261 & 0.9645& \worst 0.8032 & 0.9595& 0.8356& 0.9625 & 0.9849  & \best 1.0000   \\ \hline
\textbf{CVE-2017-5487}   & 0.9184 & 0.8835& \worst 0.8065 & 0.9044& 0.9045& 0.9347 & 0.9293  & \best 0.9901   \\ \hline
\textbf{CVE-2016-4029}   & 0.9394 & 0.9381& \worst 0.8130 & 0.9215& 0.9111& 0.9405 & 0.9495  & \best 1.0000   \\ \hline
\textbf{CVE-2023-23752}  & 0.8857 & 0.9697& 0.8130 & 0.9825& 0.8854& 0.9256 & \worst 0.7831  & \best 1.0000   \\ \hline
\textbf{CVE-2021-23132}  & 0.9179 & 0.9362& \worst 0.8032 & 0.9491& 0.9074& 0.8913 & 0.9091  & \best 0.9950   \\ \hline
\textbf{CVE-2017-8917}   & 0.9423 & 0.9238& 0.8032 & 0.9563& 0.8826& 0.9320 & \worst 0.7607  & \best 1.0000   \\ \hline
\textbf{CVE-2015-8562}   & 0.9246 & 0.9694& 0.8065 & 0.9510& 0.8973& 0.9453 & \worst 0.6832  & \best 1.0000   \\ \hline
\textbf{CVE-2019-8341}   & 0.9510 & 0.9800& \worst 0.7905 & 0.9475& 0.8978& 0.9485 & 0.9798  & \best 1.0000   \\ \hline \hline
\textbf{CVE-2019-5736}   & 0.9505 & 0.9648& \worst 0.8197 & 0.9471& 0.9482& 0.9555 & 0.9333  & \best 0.9655   \\ \hline
\textbf{CVE-2021-30465}  & 0.9223 & 0.9307& \worst 0.7937 & 0.9153& 0.9049& 0.9236 & 0.9751  & \best 1.0000   \\ \hline
\textbf{CVE-2024-21626}  & 0.9694 & 0.8155& \worst 0.8130 & 0.9513& 0.9405& 0.9633 & \best 1.0000  & \best 1.0000   \\ \hline
\textbf{CVE-2020-15257}  & 0.9417 & 0.9557& \worst 0.8130 & 0.9589& 0.9018& 0.9125 & \best 1.0000  & \best 1.0000   \\ \hline
\textbf{CVE-2022-1708}   & 0.9849 & 0.9608& \worst 0.8065 & 0.9902& 0.9809& 0.9530 & 0.9950  & \best 1.0000   \\ \hline
\textbf{CVE-2020-14386}  & 0.9192 & 0.9314& \worst 0.8368 & 0.9292& 0.9411& 0.9056 & 0.9851  & \best 1.0000   \\ \hline
\textbf{CVE-2024-1086}   & 0.9384 & 0.9474& \worst 0.8130 & 0.9431& 0.9312& 0.9662 & \best 1.0000  & \best 1.0000   \\ \hline
\textbf{CVE-2021-25742}  & 0.8900 & 0.9101& \worst 0.8130 & 0.9400& 0.9094& 0.9240 & 0.9082  & \best 0.9900   \\ \hline
\textbf{CVE-2021-25743}  & 0.9400 & 0.9347& \worst 0.8000 & 0.9206& 0.9247& 0.9404 & 0.9534  & \best 1.0000   \\ \hline
\textbf{CWE-200}  & 0.9179 & 0.9453& \worst 0.8163 & 0.9606& 0.8857& 0.9532 & 0.9851  & \best 1.0000  \\ \hline
\end{tabular}
}
\end{table}

Notably, the AUC and F1-scores for CHIDS are around 0.8 across all attacks. Through our analysis, these results are attributed to the design of CHIDS, which extracts only the numerical degree information from the system call graph for AE encoding, unlike our implementation of AE that directly encodes system calls with semantic information. Therefore, CHIDS tends to have a higher incidence of false-positive log samples under large-volume log evaluations such as cloud system logs.

\begin{tcolorbox}[size=title,opacityfill=0.1,breakable]
\noindent
\textbf{Answer to RQ1: } LAD methods can achieve an average F1-Score of 0.9 in cloud attack detection except for CHIDS, which has a relatively poor ability to detect DoS attacks. The introduced VAE model achieves nearly perfect detection results across all types of attacks.
\end{tcolorbox}

\subsection{RQ2: Effectiveness of LAD Methods under Normality Shift Scenarios}

\begin{figure}[h]
\centering
\includegraphics[width=1\linewidth]{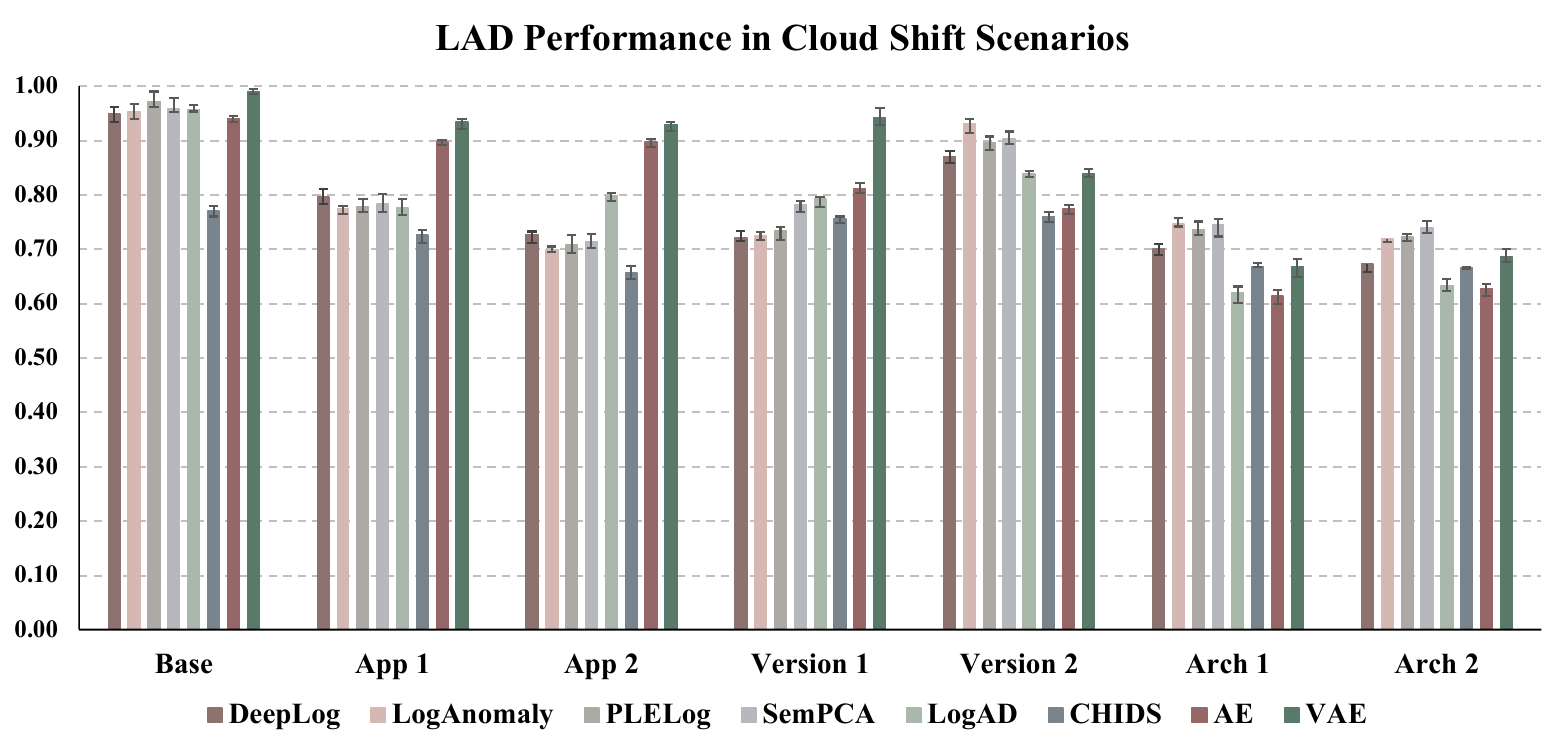}
\Description{LAD Performance in Cloud Shift Scenarios}
\caption{F1-Scores achieved by each LAD method under normality shift.}
\label{fig:rq2-result}
\end{figure}

 We then explore whether existing LAD methods can perform well in more practical and challenging normality shift scenarios. To do so, after training LAD models on the same normality distribution (Kubernetes system using \textit{containerd} and \textit{runc} deployed with \textit{WordPress} in version 6.2), we evaluate the detection capability of trained models using our collected shift datasets, CAShift. For simplicity, we represent shift logs in cloud application \textit{Jinja2} as \textit{App-1}, \textit{Joomla} as \textit{App-2}, WordPress version 4.8 as \textit{Version-1}, WordPress version 5.6 as \textit{Version-2}, cloud runtime \textit{containerd} with \textit{gVisor} as \textit{Arch-1} and cloud runtime \textit{cri-o} with \textit{runc} as \textit{Arch-2}. Due to the limited space, we have averaged the results for all evaluated attacks and presented them in Table~\ref {rq2:shift evaluation}. Besides, Figure~\ref{fig:rq2-result} depicts the F1-scores achieved by each method for better understanding. 

\begin{table}[t]
\centering
\caption{ LAD performance under normality shift scenarios. The best and worst results are highlighted by green and red background, respectively.}
\label{rq2:shift evaluation}
\resizebox{1\textwidth}{!}{
\begin{tabular}{Bc|ccccccc}
\hline
\textbf{Baseline} & \textbf{Metrics}   & \textbf{Base} & \textbf{App 1} & \textbf{App 2} & \textbf{Version 1} & \textbf{Version 2} & \textbf{Arch 1}& \textbf{Arch 2}\\ \hline
 & AUC   & 0.9460 & 0.7790 & 0.6504 & 0.6564 & 0.8576 & 0.6074 & 0.5528 \\
 & Precision & 0.9632 & 0.7000 & 0.6074 & 0.5915 & 0.8343 & 0.5663 & 0.5217 \\
 & Recall& 0.9340 & 0.9240 & 0.9040 & 0.9240 & 0.9100 & 0.9160 & 0.9480 \\
\multirow{-4}{*}{\textbf{DeepLog}}& F1-Score  & \cellcolor[HTML]{D9D9D9}0.9483 & \cellcolor[HTML]{D9D9D9}0.7965 & \cellcolor[HTML]{D9D9D9}0.7265 & \cellcolor[HTML]{F4CCCC}0.7212 & \cellcolor[HTML]{D9D9D9}0.8705 & \cellcolor[HTML]{D9D9D9}0.6998 & \cellcolor[HTML]{D9D9D9}0.6730 \\ \hline
 & AUC   & 0.9450 & 0.7268 & 0.6060 & 0.6360 & 0.9264 & 0.6948 & 0.6366 \\
 & Precision & 0.9762 & 0.6554 & 0.5565 & 0.5781 & 0.9552 & 0.6241 & 0.6038 \\
 & Recall& 0.9320 & 0.9480 & 0.9400 & 0.9680 & 0.9060 & 0.9300 & 0.8900 \\
\multirow{-4}{*}{\textbf{LogAnomaly}} & F1-Score  & \cellcolor[HTML]{D9D9D9}0.9535 & \cellcolor[HTML]{D9D9D9}0.7748 & \cellcolor[HTML]{D9D9D9}0.6991 & \cellcolor[HTML]{D9D9D9}0.7238 & \cellcolor[HTML]{D9EAD3}0.9299 & \cellcolor[HTML]{D9EAD3}0.7468 & \cellcolor[HTML]{CCCCCC}0.7194 \\ \hline
 & AUC   & 0.9650 & 0.7454 & 0.6734 & 0.6497 & 0.9156 & 0.7059 & 0.6294 \\
 & Precision   & 0.9727 & 0.6839 & 0.5568 & 0.6734 & 0.8995 & 0.6271 & 0.6134 \\
 & Recall & 0.9710 & 0.9020 & 0.9700 & 0.8050 & 0.8940 & 0.8900 & 0.8800 \\
\multirow{-4}{*}{\textbf{PLELog}}  & F1-Score& \cellcolor[HTML]{D9D9D9}0.9718 & \cellcolor[HTML]{D9D9D9}0.7780 & \cellcolor[HTML]{D9D9D9}0.7075 & \cellcolor[HTML]{D9D9D9}0.7333 & \cellcolor[HTML]{D9D9D9}0.8967 & \cellcolor[HTML]{D9D9D9}0.7358 & \cellcolor[HTML]{D9D9D9}0.7229 \\ \hline
 & AUC   & 0.9533 & 0.7941 & 0.6778 & 0.6742 & 0.8906 & 0.6840 & 0.6106 \\
 & Precision & 0.9668 & 0.6872 & 0.5614 & 0.7184 & 0.8873 & 0.6234 & 0.6278 \\
 & Recall& 0.9490 & 0.9105 & 0.9798 & 0.8564 & 0.9187 & 0.9269 & 0.8989 \\
\multirow{-4}{*}{\textbf{SemPCA}} & F1-Score  & \cellcolor[HTML]{D9D9D9}0.9578 & \cellcolor[HTML]{D9D9D9}0.7832 & \cellcolor[HTML]{D9D9D9}0.7138 & \cellcolor[HTML]{D9D9D9}0.7814 & \cellcolor[HTML]{D9D9D9}0.9027 & \cellcolor[HTML]{D9D9D9}0.7454 & \cellcolor[HTML]{D9EAD3}0.7393 \\ \hline
 & AUC   & 0.9523 & 0.7357 & 0.6962 & 0.6978 & 0.9224 & 0.7139 & 0.6573 \\
 & Precision & 0.9782 & 0.6573 & 0.6896 & 0.6734 & 0.7497 & 0.4602 & 0.4795 \\
 & Recall& 0.9353 & 0.9467 & 0.9491 & 0.9612 & 0.9504 & 0.9488 & 0.9316 \\
\multirow{-4}{*}{\textbf{LogAD}} & F1-Score  & \cellcolor[HTML]{D9D9D9}0.9563 & \cellcolor[HTML]{D9D9D9}0.7759 & \cellcolor[HTML]{D9D9D9}0.7988 & \cellcolor[HTML]{D9D9D9}0.7920 & \cellcolor[HTML]{D9D9D9}0.8382 & \cellcolor[HTML]{D9D9D9}0.6198 & \cellcolor[HTML]{D9D9D9}0.6331 \\ \hline
 & AUC   & 0.7082 & 0.6360 & 0.5604 & 0.6726 & 0.6690 & 0.5060 & 0.4952 \\
 & Precision & 0.6275 & 0.5708 & 0.4911 & 0.6101 & 0.6178 & 0.5017 & 0.4992 \\
 & Recall& 0.9960 & 0.9960 & 0.9900 & 0.9920 & 0.9880 & 1.0000 & 1.0000 \\
\multirow{-4}{*}{\textbf{CHIDS}}  & F1-Score  & \cellcolor[HTML]{E6B8AF}0.7699 & \cellcolor[HTML]{E6B8AF}0.7256 & \cellcolor[HTML]{E6B8AF}0.6565 & \cellcolor[HTML]{D9D9D9}0.7555 & \cellcolor[HTML]{E6B8AF}0.7602 & \cellcolor[HTML]{D9D9D9}0.6682 & \cellcolor[HTML]{D9D9D9}0.6660 \\ \hline
 & AUC   & 0.9260 & 0.9890 & 0.9921 & 0.9033 & 0.8956 & 0.8802 & 0.7078 \\
 & Precision & 0.9943 & 0.9655 & 0.8780 & 0.8261 & 0.6581 & 0.5576 & 0.5439 \\
 & Recall& 0.8930 & 0.8375 & 0.9175 & 0.7990 & 0.9375 & 0.6831 & 0.7375 \\
\multirow{-4}{*}{\textbf{AE}} & F1-Score  & \cellcolor[HTML]{D9D9D9}0.9409 & \cellcolor[HTML]{D9D9D9}0.8970 & \cellcolor[HTML]{D9D9D9}0.8973 & \cellcolor[HTML]{D9D9D9}0.8123 & \cellcolor[HTML]{D9D9D9}0.7733 & \cellcolor[HTML]{F4CCCC}0.6140 & \cellcolor[HTML]{F4CCCC}0.6261 \\ \hline
 & AUC   & 0.9979 & 0.9929 & 0.9950 & 0.9942 & 0.9781 & 0.7384 & 0.7011 \\
 & Precision & 0.9976 & 0.8772 & 0.8672 & 0.8911 & 0.7240 & 0.5020 & 0.5228 \\
 & Recall& 0.9820 & 0.9995 & 0.9995 & 0.9995 & 0.9995 & 0.9995 & 0.9995 \\
\multirow{-4}{*}{\textbf{VAE}}& F1-Score  & \cellcolor[HTML]{D9EAD3}0.9897 & \cellcolor[HTML]{D9EAD3}0.9344 & \cellcolor[HTML]{D9EAD3}0.9287 & \cellcolor[HTML]{D9EAD3}0.9422 & \cellcolor[HTML]{D9D9D9}0.8397 & \cellcolor[HTML]{D9D9D9}0.6683 & \cellcolor[HTML]{D9D9D9}0.6865 \\ \hline
\end{tabular}
}
\end{table}

\fish{From the results, we can see that normality shift highly affects the effectiveness of LAD methods, leading to varying degrees of performance decrease across different scenarios. We note that most of the existing LAD methods (e.g., \textit{DeepLog} and \textit{LogAnomaly}) experienced significant reductions (averaging around 20\%) in F1-score across all shift scenarios except for scenario \textit{Version-2}. For harsh shift scenarios such as \textit{Arch-1} and \textit{Arch-2}, methods in couples with log term frequency statistics, such as \textit{LogAnomaly} and \textit{SemPCA}, demonstrate more robust detection capabilities compared to other baselines. For the advanced method \textit{LogAD}, the database-assisted design leads to a greater decrease in LAD performance with many false positives.
Surprisingly, the introduced VAE model achieves almost perfect detection results under application and version shift scenarios, demonstrating that the near-perfect detection results in Section~\ref{sec:rq1} are not attributable to overfitting. However, VAE is also significantly influenced by the cloud architecture shift with many false positives, which is a unique shift type in cloud systems compared to traditional systems. The result indicates that VAE is unable to differentiate between severely shifted normal logs and attack logs.}

\fish{To provide a robust statistical foundation for assessing how shift logs and their characteristics influence LAD models, we conduct an independent samples \textit{t-test}. This analysis evaluates the statistical significance of differences between in-distribution and six-shift log data. The p-values for the F1-scores of LAD models in each shift scenario are 1.1E-03, 1.0E-03, 1.3E-03, 3.9E-03, 4.7E-05, and 2.1E-05, respectively. These results indicate distribution shift significantly affects each method.}

\begin{tcolorbox}[size=title,opacityfill=0.1,breakable]
\noindent
\textbf{Answer to RQ2:} All LAD methods are impacted by normality shift, which leads to decreased detection performance by up to 34\% in F1-Score. Specifically, VAE has the best and most robust performance against most shift scenarios, while having a significant decrease in performance during cloud architecture shifts.
\end{tcolorbox}

\subsection{RQ3: Effectiveness of Shift Adaption}
\label{sec:rq3}
\begin{figure}[h]
\centering

\begin{subfigure}[]{0.495\textwidth}
\centering
\includegraphics[width=\textwidth]{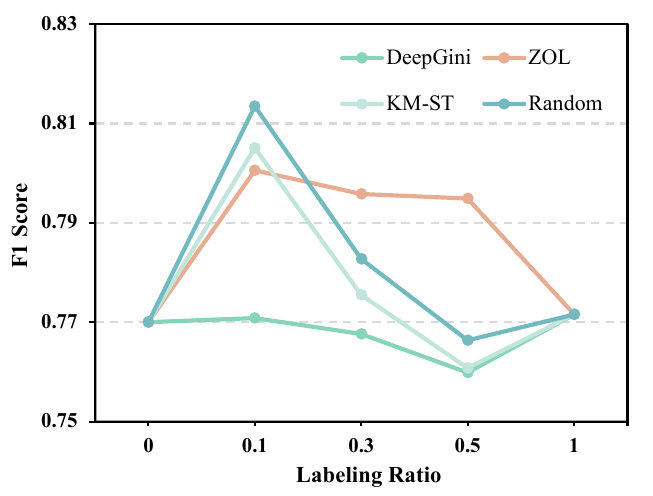}
\caption{Average performance change of AE model}
\label{fig:rq3_ae}
\end{subfigure}
\hfill
\begin{subfigure}[]{0.495\textwidth}
\centering
\includegraphics[width=\textwidth]{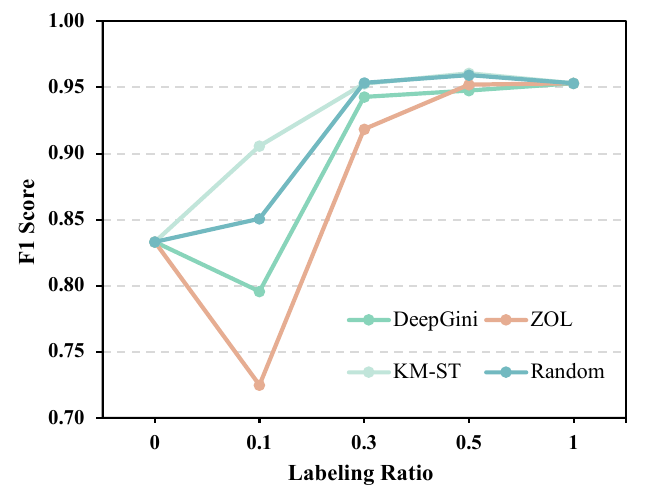}
\caption{Average performance change of VAE model}
\label{fig:ra3_vae}
\end{subfigure}

\caption{The average F1-Score changes during continuous learning.}
\Description{Average F1-Score for RQ3}
\label{fig:rq3}
\end{figure}

Our evaluation in RQ2 reveals that normality shift can introduce performance drops to LAD methods, therefore, we further investigate whether continuous learning-based shift adaptation can help mitigate such negative impacts. Here, we only consider AE and VAE models as they are relatively better methods under various distribution shifts. We apply four continuous learning methods in updating the LAD models under shift scenarios. \fish{We follow previous works to explore hyper-parameters~\cite{le2022log} to determine the optimal performance of LAD under different retraining budgets~\cite{han2023anomaly,hu2022empirical}.} Table~\ref{tab: rq3} summarizes the F1-Scores for each baseline after retraining, and  Figure~\ref{fig:rq3} depicts the average performance changes of retrained LAD models across different shift scenarios.

\begin{table}[h]
\centering
\caption{Results of continuous learning for LAD shift adaptation. The best and worst results are highlighted by the green and red background, respectively.} 
\resizebox{1\textwidth}{!}{
\begin{tabular}{|B|B|c|c|c|c|c|c|c|}
\hline
\textbf{Method}  & \textbf{Baseline} & \textbf{Percentage} & \textbf{App 1}  & \textbf{App 2}  & \textbf{Version 1}    & \textbf{Version 2}    & \textbf{Arch 1} & \textbf{Arch 2} \\ \hline
     & \textbf{Shift}    & 0 & 0.8970 & 0.8973 & 0.8123 & 0.7733 & 0.6140 & 0.6261 \\ \cline{2-9} 
     & & 0.1  & 0.8454 & 0.8540 & 0.8784 & \cellcolor[HTML]{F4CCCC}0.7491 & 0.6479 & 0.6502 \\ \cline{3-9} 
     & & 0.3  & 0.8140 & 0.8243 & 0.8720 & 0.8038 & \cellcolor[HTML]{F4CCCC}0.6449 & 0.6464 \\ \cline{3-9} 
     & \multirow{-3}{*}{\textbf{DeepGini}} & 0.5  & \cellcolor[HTML]{F4CCCC}0.8119 & 0.8150 & 0.8624 & 0.7937 & 0.6484 & \cellcolor[HTML]{F4CCCC}0.6279 \\ \cline{2-9} 
     & & 0.1  & 0.8287 & 0.8271 & 0.8815 & 0.7565 & 0.8579 & 0.6520 \\ \cline{3-9} 
     & & 0.3  & 0.8198 & 0.8184 & 0.8582 & 0.7526 & 0.8784 & 0.6475 \\ \cline{3-9} 
     & \multirow{-3}{*}{\textbf{ZOL}}& 0.5  & 0.8131 & \cellcolor[HTML]{F4CCCC}0.8113 & 0.8560 & 0.7585 & \cellcolor[HTML]{D9EAD3}0.8857 & 0.6448 \\ \cline{2-9} 
     & & 0.1  & 0.8551 & \cellcolor[HTML]{D9EAD3}0.8544 & \cellcolor[HTML]{D9EAD3}0.9015 & 0.7507 & 0.8041 & 0.6644 \\ \cline{3-9} 
     & & 0.3  & 0.8303 & 0.8287 & 0.8734 & 0.7661 & 0.7083 & 0.6464 \\ \cline{3-9} 
     & \multirow{-3}{*}{\textbf{KM-ST}}    & 0.5  & 0.8150 & 0.8201 & 0.8714 & 0.7644 & 0.6487 & 0.6448 \\ \cline{2-9} 
     & & 0.1  & \cellcolor[HTML]{D9EAD3}0.8561 & 0.8513 & 0.8994 & 0.8190 & 0.7867 & \cellcolor[HTML]{D9EAD3}0.6682 \\ \cline{3-9} 
     & & 0.3  & 0.8175 & 0.8333 & 0.8755 & 0.8210 & 0.7024 & 0.6466 \\ \cline{3-9} 
     & \multirow{-3}{*}{\textbf{Random}}   & 0.5  & 0.8133 & 0.8180 & \cellcolor[HTML]{F4CCCC}0.8559 & 0.8177 & 0.6487 & 0.6448 \\ \cline{2-9}
\multirow{-14}{*}{\textbf{AE}}  & \textbf{All}& 1.0  & 0.8156 & 0.8147 & 0.8585 & \cellcolor[HTML]{D9EAD3}0.8266 & 0.6619 & 0.6524 \\ \hline \hline
     & \textbf{Shift}    & 0 & 0.9344 & 0.9287 & 0.9422 & 0.8397 & 0.6683 & 0.6865 \\ \cline{2-9} 
     & & 0.1  & \cellcolor[HTML]{F4CCCC}0.5726 & 0.8823 & 0.9548 & \cellcolor[HTML]{F4CCCC}0.7510 & 0.8768 & 0.7374 \\ \cline{3-9} 
     & & 0.3  & 0.9672 & 0.9511 & 0.9660 & 0.9562 & 0.9260 & 0.8908 \\ \cline{3-9} 
     & \multirow{-3}{*}{\textbf{DeepGini}} & 0.5  & 0.9868 & 0.9707 & 0.9634 & 0.9614 & 0.9100 & 0.8936 \\ \cline{2-9} 
     & & 0.1  & 0.6669 & \cellcolor[HTML]{F4CCCC}0.6665 & 0.9554 & 0.7726 & \cellcolor[HTML]{F4CCCC}0.6999 & \cellcolor[HTML]{F4CCCC}0.5878 \\ \cline{3-9} 
     & & 0.3  & 0.9736 & 0.9328 & 0.9641 & 0.9634 & 0.8146 & 0.8627 \\ \cline{3-9} 
     & \multirow{-3}{*}{\textbf{ZOL}}& 0.5  & \cellcolor[HTML]{D9EAD3}0.9922 & 0.9482 & 0.9619 & 0.9552 & 0.9175 & \cellcolor[HTML]{D9EAD3}0.9373 \\ \cline{2-9} 
     & & 0.1  & 0.9021 & 0.8212 & 0.9661 & 0.8819 & 0.9373 & 0.9249 \\ \cline{3-9} 
     & & 0.3  & 0.9527 & 0.9744 & 0.9604 & 0.9622 & 0.9381 & 0.9296 \\ \cline{3-9} 
     & \multirow{-3}{*}{\textbf{KM-ST}}    & 0.5  & 0.9730 & 0.9845 & 0.9646 & 0.9695 & 0.9378 & 0.9345 \\ \cline{2-9} 
     & & 0.1  & 0.6686 & 0.8167 & \cellcolor[HTML]{F4CCCC}0.9301 & 0.8845 & 0.9280 & 0.8764 \\ \cline{3-9} 
     & & 0.3  & 0.9581 & 0.9806 & \cellcolor[HTML]{D9EAD3}0.9664 & 0.9689 & 0.9290 & 0.9172 \\ \cline{3-9} 
     & \multirow{-3}{*}{\textbf{Random}}   & 0.5  & 0.9839 & \cellcolor[HTML]{D9EAD3}0.9924 & 0.9619 & \cellcolor[HTML]{D9EAD3}0.9743 & \cellcolor[HTML]{D9EAD3}0.9404 & 0.9029 \\ \cline{2-9} 
\multirow{-14}{*}{\textbf{VAE}} & \textbf{All}& 1.0  & 0.9763 & 0.9773 & 0.9614 & 0.9742 & 0.9342 & 0.8965 \\ \hline
\end{tabular}
}
\label{tab: rq3}
\end{table}

\fish{The results demonstrate that retraining with a portion of new data can indeed enhance the performance of LAD models on the shifted test set. In all cases, the LAD models with the best results are models after shift adaptation. However, we notice that retraining may reduce the performance of LAD models, e.g., DeepGini with a ratio of 0.1. Thus, selecting an appropriate number of logs and choosing the right log selection strategy are both crucial for LAD shift adaptation. More specifically, our results indicate that for the AE model, the \textit{ZOL} selection strategy generally outperforms the other three methods, with an average improvement of 9.88\%. Surprisingly, retraining with more data (e.g., with all available data) tends to decrease the performance of the AE model. We conjecture that AE models overfit the shifted distribution and therefore, have low performance after retraining.} 

\fish{For the VAE model, selecting a relatively larger amount of data for retraining leads to better performance improvements, with a maximum increase of 27\% in the F1-Score.  As the VAE possesses a strong capability to learn distribution information, insufficient retraining data can cause confusion between previously learned patterns and the new distribution, resulting in a significant drop in performance, as illustrated in Figure~\ref{fig:ra3_vae}. However, utilizing the entire retraining data budget can lead to diminished performance due to catastrophic forgetting of previously learned distributions. Therefore, it shows that, unlike the AE, the VAE may require more shift information to better learn from and successfully reconstruct shifted normal behavior.}


\noindent
\fish{\textbf{Labeling and Computational Cost.} Due to the data-centric nature of the LAD task, the performance of LAD models after continuous learning can vary significantly across different labeling budgets. As continuous learning methods require labeled data for retraining, the efficiency of labeling and retraining warrants discussion. Specifically, the human labeling overhead is difficult to quantify due to multiple reasons such as the experience of annotators and the complexity of tasks. Our evaluation reveals that the computational cost of model training is negligible compared to the human effort required for labeling. During our experiment, processing embeddings for the dataset of 150 million log entries takes an average of two hours, and training the models (AE and VAE) on one NVIDIA L40 GPU takes just one minute.}

\begin{tcolorbox}[size=title,opacityfill=0.1,breakable]
\noindent
\textbf{Answer to RQ3: } Continuous learning can enhance the performance of LAD models in shift environments but not always. The selection of continuous learning methods and the labeling budget highly affects the performance of the produced LAD models. Among our studied continuous learning methods, \textit{ZOL} is relatively better than other continuous learning methods.
\end{tcolorbox}

\section{Discussion}

\subsection{Hybrid Shift Scenarios and Multi-Vector Attack Analysis}

\fish{In practice, the evolution of cloud business operations can lead to hybrid normality shifts, which involve more than one shift type. For example, both the deployed application versions and the cloud system runtimes may change after the system migration, while standalone application changes or version updates generally do not result in hybrid shift behaviors. Conversely, multiple attacks are more commonly observed~\cite{cloudflare-report}, as attackers often employ various techniques simultaneously to exploit target systems effectively, aiming to maximize the attack efficiency. Therefore, we evaluate the performance of LAD under hybrid shift scenarios and multi-vector attacks.} 

\noindent
\textbf{Hybrid Shift Scenarios.} \fish{We select our collected logs containing multiple shift types and evaluate the performance of AE and VAE models accordingly, with detailed results presented in Table~\ref{table: hybrid shift}. We find that hybrid shift logs are more difficult to distinguish from attacks for LAD models. Both AE and VAE exhibit increased reconstruction loss for logs collected from environments with multiple shifts, leading to more false positives and decreased performance for LAD models.}

\begin{table}[!t]
\centering
\caption{Performance of LAD under hybrid shift scenarios.}
\label{table: hybrid shift}
\resizebox{1\textwidth}{!}{
\begin{tabular}{|c|c|c|c|c|c|c|c|c|}
\hline
    &    & \textbf{Base}  & \textbf{App 1}  & \textbf{Version 1}  & \textbf{Arch 1} & \textbf{Version + Arch}   & \textbf{Version + Arch}   & \textbf{App + Arch} \\ \cline{3-9} 
\multirow{-2}{*}{\textbf{LAD Model}} & \multirow{-2}{*}{\textbf{Metrics}}  & WordPress-6-runC & Joomla-3-runC    & WordPress-5-runC & WordPress-6-crio & WordPress-5-crio & WordPress-4-crio & Joomla-3-crio    \\ \hline
    & AUC& 0.9260 & 0.9890 & 0.9033 & 0.8802 & 0.8668 & 0.8680 & 0.8678 \\ \cline{2-9} 
    & Precision& 0.9943 & 0.9655 & 0.8261 & 0.5576 & 0.4719 & 0.4789 & 0.4034 \\ \cline{2-9} 
    & Recall & 0.8930 & 0.8375 & 0.7990 & 0.6831 & 0.7600 & 0.7600 & 0.7600 \\ \cline{2-9} 
\multirow{-4}{*}{\textbf{AE}} & \cellcolor[HTML]{EFEFEF}F1-Score & \cellcolor[HTML]{EFEFEF}0.9409 & \cellcolor[HTML]{EFEFEF}0.8970 & \cellcolor[HTML]{EFEFEF}0.8123 & \cellcolor[HTML]{EFEFEF}0.6140 & \cellcolor[HTML]{EFEFEF}0.5823 & \cellcolor[HTML]{EFEFEF}0.5878 & \cellcolor[HTML]{EFEFEF}0.5270 \\ \hline
    & AUC& 0.9979 & 0.9929 & 0.9942 & 0.7384 & 0.6766 & 0.6428 & 0.6435 \\ \cline{2-9} 
    & Precision& 0.9976 & 0.8772 & 0.8911 & 0.5020 & 0.4842 & 0.4633 & 0.4236 \\ \cline{2-9} 
    & Recall & 0.9820 & 0.9995 & 0.9995 & 0.9995 & 0.9850 & 0.9850 & 0.9850 \\ \cline{2-9} 
\multirow{-4}{*}{\textbf{VAE}}& \cellcolor[HTML]{EFEFEF}F1-Score & \cellcolor[HTML]{EFEFEF}0.9897 & \cellcolor[HTML]{EFEFEF}0.9344 & \cellcolor[HTML]{EFEFEF}0.9422 & \cellcolor[HTML]{EFEFEF}0.6683 & \cellcolor[HTML]{EFEFEF}0.6492 & \cellcolor[HTML]{EFEFEF}0.6302 & \cellcolor[HTML]{EFEFEF}0.5924 \\ \hline
\end{tabular}
}
\end{table}

\noindent
\textbf{Multi-Vector Attack Scenarios}. We collect logs from the cloud system while launching multiple simultaneous attacks on different cloud components to evaluate LAD models in detecting multi-vector attacks. The detailed attack information can be found on our website~\cite{website}. Experimental results reveal that these attacks can all be successfully detected by the LAD models. However, the anomaly scores of these attacks show that multi-vector attacks present more distinct anomaly characteristics. Consequently, these characteristics make it easier to differentiate from normal logs.



\subsection{Implications} 

Based on our evaluation results and findings, we identify several key implications that could serve as promising guidance for future research on log-based attack detection.

\begin{itemize}[left=0.1cm]
    \item \textbf{Emphasize the importance of normality shift in LAD.} Our experimental results demonstrate that even the best VAE model, which has almost perfect detection performance on most shift scenarios, experiences a significant drop in performance under a specific type of shift, i.e., cloud architecture shift. Such drops could negatively affect the practical usage of LAD methods in real-world scenarios. \fish{Therefore, although numerous arts are proposed to improve the performance of LAD under in-distribution environment detection, researchers should pay more attention to exploring the robustness of these models under normality shift scenarios. The cloud system maintainers should collaborate closely with tenants to maintain an active LAD updating scheme and consider applying existing distribution shift detector tools~\cite{devries2018learning,hendrycks2018deep}.}

    \item \textbf{Carefully select shift adaptation methods for different LAD approaches.} Shift adaptation is necessary due to the significant performance drops of LAD methods under normality shift scenarios. \fish{Our evaluation highlights the overall positive impact of continuous learning methods on improving LAD performance in various shift scenarios (e.g., different applications and cloud systems). However, the selection of continuous learning algorithms and the proportion of labeled post-shift data used for retraining remain crucial. It involves trade-offs between labeling efficiency and model effectiveness. As demonstrated by our experimental results in Section~\ref{sec:rq3}, choosing sub-optimal adaptation methods cannot only fail to enhance the performance of LAD but may also worsen its capabilities. Therefore, developers should iteratively increase their budgets and check the performance of LAD models frequently. Our results show that identical shifted samples should be prioritized (i.e., ZOL) for selection in LAD shift adaptation.}
    
    \item \textbf{Reconsider log distribution in actual application scenarios.} When adopting continuous learning to enhance LAD performance, the learned knowledge from new distributions could affect the detection ability for already learned data with old distributions. As discussed in prior works~\cite{han2023anomaly,du2019lifelong}, the phenomenon of knowledge learned previously being erased when learning new samples is termed as \textit{catastrophic forgetting}. \fish{This problem needs to be considered during updating LAD models. For example, if certain application functionalities no longer exist in new scenarios, they are regarded as outdated data. In such cases, shift adaptation efforts should focus solely on the new data distributions. Alternatively, developers may need to adapt to new distributions while also ensuring that LAD models retain the knowledge of older distributions.}
    
\end{itemize}

\subsection{Future Research Directions}


Based on our findings, we identify the following research challenges and provide potential solutions.
\begin{itemize}[left=0.1cm]
    \item \textit{Diverse data preparation.} Our study underscores the need to prepare more diverse datasets to evaluate the capability of LAD applications across various domains, particularly those featuring different normality shift scenarios. LAD models that achieve promising detection results on a single distribution do not guarantee robust performance across other distribution characteristics.

    \item  \textit{Multi-modal consideration.} In cloud attack detection scenarios, there are various types of logs available besides system call logs, including network traffic information, application-based cloud monitors such as Falco~\cite{falco}, and application-specific logs such as Java runtime logs. Therefore, in addition to expanding log datasets from diverse data sources, considering multi-modal methods to enhance LAD performance in cloud attack scenarios could also be beneficial. This approach could improve the current LAD models' ability to learn and detect attack characteristics that are difficult to discern solely from system call logs.
        
    \item \textit{Continuous learning methods enhancement.} While label-efficient continuous learning methods have shown potential for mitigating performance degradation in existing LAD models under shift scenarios, significant room for improvement remains, particularly in cases of substantial shifts where different models exhibit varied performance changes. Adapting different algorithms to optimally train models on new distribution log patterns could also be a direction for boosting LAD shift adaptation. For example, integrating methods like contrasting learning algorithms~\cite{wang2022robust} in continuous learning methods.
    
    \item \textit{Other shift adaptation methods consideration.} To enhance LAD model performance in shift scenarios, other training-less methods~\cite{el2024replicawatcher,pan2024raglog} can also improve LAD performance, such as retrieval augmentation generation (RAG). These methods utilize fewer labeled true positive logs as references to learn post-shift log characteristics without retraining. However, these methods still struggle with generalization and require specific retrieval sample preparation for each shift scenario. Therefore, constructing RAG retrieval for LAD is also a promising research direction.
    
\end{itemize}

\subsection{Threat to Validity}


The internal threats to validity are the implementations of models and baselines. During our evaluation, we unify the result-reporting logic of existing baselines to ensure fair baseline comparison. We prioritize using publicly available implementations for the baselines. For those without open-sourced implementations, we implement the models based on detailed descriptions in the respective papers. We maintain consistency in hyperparameters for these baselines to mitigate potential threats. Specifically, we conduct evaluations on existing datasets (including \textit{CB-DS}, \textit{LogHub}) first to verify correctness, and then proceed with our benchmark.


The external threats come from the collected datasets, used models, and baselines. During the construction of dataset \dataset, as we cannot collect all attack behaviors from cloud systems, we collect 20 of the most representative attacks, which consider both the severity and diversity of our attack scenarios in \dataset to evaluate the performance of LAD methods. We also include the three most common shift types in cloud systems in our collected dataset. In the future, we plan to continuously maintain our dataset and add more attacks that consider additional cloud shift types to better evaluate existing LAD methods. For the evaluation of baselines, we use eight representative LAD models. We exclude supervised models from our evaluation due to the high cost of labeling inherent in cloud log characteristics, rendering supervised methods impractical. In future work, we plan to include more models to demonstrate the generalizability of our conclusions.



\section{Conclusion}

In this paper, we constructed a shift-aware cloud attack dataset, \dataset for assisting the security assurance of cloud systems. Using \dataset, we conducted an extensive study to explore the capabilities of LAD methods in detecting cloud attacks under normality shift scenarios. We also examined the usefulness of current continuous learning methods for LAD shift adaptation. We revealed that existing LAD methods are significantly impacted by normality shifts, and uncertainty-based continuous learning strategies are promising for shift adaptation. Based on our findings, we provided potential research directions for future studies on LAD in shift environments. We believe our work lays the groundwork for future research in the important area of cloud attack detection.

\section{Data Availability}
The \dataset dataset together with the source code for experiment evaluation involved in this work can be found at our website~\cite{website}.


\section*{Acknowledgment}

This research is partially supported by the Lee Kong Chian Fellowship, the National Research Foundation, Singapore, and the Cyber Security Agency under its National Cybersecurity R\&D Programme (NCRP25-P04-TAICeN). Any opinions, findings and conclusions or recommendations expressed in this material are those of the author(s) and do not reflect the views of National Research Foundation, Singapore and Cyber Security Agency of Singapore.


\bibliographystyle{ACM-Reference-Format}
\bibliography{fse2025}

\appendix

\end{document}